\documentclass[onecolumn,superscriptaddress,floatfix,preprintnumbers, nofootinbib,hyperref]{revtex4} 
\usepackage{graphicx}%  Default Latex 2eps package for embedding figures
\usepackage{rotating}
\usepackage{epsfig}
\usepackage{bm, amsmath, bigints, amsfonts}
\usepackage[usenames,dvipsnames]{xcolor}

\bibpunct{[}{]}{,}{n}{}{,}

\begin{document}

\title{Fast flavor conversions of supernova neutrinos: \\ Classifying  instabilities via dispersion relations}

\author{Francesco\,Capozzi} 
\affiliation{Department of Physics, Ohio State University, Columbus, OH 43210, USA}

\author{Basudeb\,Dasgupta} 
\affiliation{Tata Institute of Fundamental Research,
      Homi Bhabha Road, Mumbai, 400005, India.}

\author{Eligio\,Lisi} 
\affiliation{Istituto Nazionale di Fisica Nucleare - Sezione di Bari, Via Orabona 4, 70126 Bari, Italy}

\author{Antonio\,Marrone} 
\affiliation{Dipartimento Interateneo di Fisica ``Michelangelo Merlin'', Via Amendola 173, 70126 Bari, Italy}
\affiliation{Istituto Nazionale di Fisica Nucleare - Sezione di Bari, Via Orabona 4, 70126 Bari, Italy}

\author{Alessandro\,Mirizzi} 
\affiliation{Dipartimento Interateneo di Fisica ``Michelangelo Merlin'', Via Amendola 173, 70126 Bari, Italy}
\affiliation{Istituto Nazionale di Fisica Nucleare - Sezione di Bari, Via Orabona 4, 70126 Bari, Italy}

\date{\today}
%=============================================================================

\begin{abstract}
Supernova neutrinos can exhibit a rich variety of flavor conversion mechanisms. In particular, they
can experience 
``fast'' self-induced flavor conversions almost immediately above the core. 
Very recently, a novel method has been proposed to investigate these phenomena, in terms  
of the dispersion relation for the complex frequency and wave number $(\omega,\, {k})$ of 
disturbances in the mean field of the $\nu_e\nu_x$ flavor coherence. We discuss a systematic approach to such 
instabilities, originally developed in the context of plasma physics,
and based of the time-asymptotic behavior of the Green's function of the system.
Instabilities are typically seen to emerge for complex  
 ${\omega}$, and can be further characterized  as convective 
 (moving away faster than they spread) and absolute (growing locally), depending on ${k}$-dependent features.
Stable cases emerge when $k$ (but not $\omega$) is complex, leading to disturbances damped in space, or when both
$\omega$ and $k$ are real, corresponding to complete stability.  
The analytical classification of both unstable and stable modes leads not only to qualitative
insights about their features but also to quantitative predictions about the growth rates of instabilities.
Representative numerical solutions are discussed in a simple two-beam model of interacting neutrinos. 
As an application, we argue that supernova and binary neutron star mergers  exhibiting a ``crossing'' in the electron lepton number would lead
to an absolute instability in the flavor content of the neutrino gas. 
\end{abstract}

\maketitle

\section{Introduction}

Flavor conversions of supernova (SN) neutrinos are a field of intense investigation. 
Conversions in the deepest SN regions would have a potential impact on the explosion dynamics (e.g., the revival of the shock-wave) and on the nucleosynthesis of heavy elements in the stellar matter.  Flavor transition effects, occurring during neutrino propagation from inner to outer SN regions, would also imprint peculiar features on observable spectra in the next Galactic SN neutrino burst  (see\,\cite{Mirizzi:2015eza} for a recent review).
It is expected that different regimes of flavor conversions $\nu_e\to\nu_x$ ($x=\mu,\,\tau$) would be encountered by SN neutrino with energy $E$ and squared mass difference $\Delta m^2$ while they propagate through background fermions with density $n$.
Early studies focussed on the so-called Mikheyev-Smirnov-Wolfenstein (MSW) matter effects\,\cite{Wolfenstein:1977ue,Mikheev:1986gs}, occurring when the neutrino oscillation frequency in vacuum $ \omega_{\rm vac}=\Delta m^2/(2E)$ is comparable to the matter potential $\lambda = \sqrt{2} G_F n_e$ induced by the
net electron density $n_e$. This condition, tipically satisfied at ``large'' distances $r \sim {\mathcal O} (10^4)$\,km  from the SN core, can produce 
resonant flavor conversions\,\cite{Dighe:1999bi} that carry information 
on the SN matter profile, including features related to the
shock wave\,\cite{Schirato:2002tg,Fogli:2003dw,Fogli:2004ff,Tomas:2004gr} and turbulence\,\mbox{\cite{Dasgupta:2005wn,Kneller:2010sc, Borriello:2013tha}}.
Later it was realized that also the SN neutrino density $n_\nu$ can be so high to provide a background medium inducing neutrino flavor change\,\cite{Pantaleone:1992xh,Pantaleone:1994ns}. When the neutrino-neutrino interaction potential $\mu \sim \sqrt{2}G_F n_\nu$ is dominant, the neutrino flavor evolution becomes nonlinear\,\cite{Pantaleone:1992eq}.  
First large-scale numerical studies of these effects\,\cite{Duan:2006an,Fogli:2007bk}, supported by analytical insights\,\cite{Hannestad:2006nj}, stimulated a torrent of  still ongoing activities. Through these studies it was realized that neutrino-neutrino interactions can lead to ``self-induced'' or ``collective'' flavor 
conversions at typical radii $r  \sim {\mathcal O}(10^2-10^3)$\,km 
and frequencies  $\sim(\omega_{\rm vac} \mu)^{1/2}$, leading to spectral swaps and splits\,\cite{Fogli:2007bk,Dasgupta:2009mg,Gava:2009pj,Duan:2010bg,Dasgupta:2010cd}.
   The robustness of these effects has been recently questioned, since
relaxing some of the symmetry assumptions imposed on simplified emission models 
\cite{Raffelt:2013rqa,Mirizzi:2013rla,Mangano:2014zda,Duan:2014gfa,Mirizzi:2015fva,Chakraborty:2015tfa,Abbar:2015fwa,Dasgupta:2015iia,Capozzi:2016oyk}
can lead to dramatic changes and even to flavor decoherence (see\,\cite{Chakraborty:2016yeg} for a recent review).
In any case, it was implicitly assumed in the literature that all these effects would occur relatively far from the neutrino emission region, being suppressed by 
dominant matter terms a closer radii\,\cite{EstebanPretel:2007ec,Chakraborty:2011nf,Chakraborty:2011gd,Chakraborty:2014nma}.

In contrast, little attention was devoted to the possibility of fast flavor conversions (with frequencies as large $\sim \mu$) at much shorter radii (with $r \sim {\mathcal O}(1)$\,m from the neutrino sphere) highlighted in \cite{Sawyer:2005jk} and developed in simple toy models\,\cite{Sawyer:2005jk,Sawyer:2008zs,Sawyer:2015dsa}.  Surprisingly, these conversions would be so fast ($\mu\gg \omega_{\rm vac}$) {that they would} wipe out any dependence on the vacuum mass-mixing parameters whose role is just to provide a ``seed'' for the self-induced flavor dynamics.   
The key  insight \cite{Sawyer:2005jk,Sawyer:2008zs,Sawyer:2015dsa} is that near the neutrino decoupling region the angular distributions of the different  neutrino species are rather different and this would cause a speed-up of the flavor conversions. 
Notably, non-electron species $\nu_x$ would decouple earlier (and thus would be more forward peaked) 
than the electron species $\nu_e$ and $\bar{\nu}_e$ and. In turn,  due to the  neutron richness of stellar matter, the $\bar\nu_e$ would decouple earlier (and thus would be more forward peaked) than ${\nu}_e$.
On the other hand, all other studies on self-induced flavor conversions had assumed the same (half-isotropic) angular distribution for any species. This approximation is reasonable for slow flavor conversions far from the neutrinosphere% 
%----------------
\footnote{See however\,\protect\cite{Mirizzi:2011tu,Mirizzi:2012wp} for a discussion of the impact of non-trivial angular distributions on slow self-induced conversions.}
but misses a crucial ingredient of fast conversions, namely, backward-going modes.
Recently, fast flavor conversions have been independently studied by some authors\,\cite{Chakraborty:2016lct,Dasgupta:2016dbv} finding that these can be possible near the SN core if the neutrino flux ratios and angular asymmetries produce a
 \emph{crossing} between the zenith-angle spectra of $\nu_e$ and ${\bar\nu}_e$. In particular, the presence of  neutrinos traveling towards the core can make these fast conversions more generic\,\cite{Dasgupta:2016dbv}. 
Fast self-induced flavor conversions are still in an exploratory phase. However, their characterization is crucial in order to have a reliable description of the neutrino flavor evolution during a stellar collapse.

In the emerging literature about fast flavor conversions, a novel 
and very interesting 
approach to study these effects was recently proposed in\,\cite{Izaguirre:2016gsx}. This approach is based on the \emph{dispersion relation} for the frequency and wavenumber 
$(\omega, {\bf k})$ in the mean field of $\nu_e\nu_x$ coherence, which is 
essentially the off-diagonal element of the neutrino density matrix $\varrho({\bf p}, {\bf x},t)$, that we will call $S$ in the following.
One looks for solutions of
the linearized equations for the flavor evolution in the
form 
%....................................
\begin{equation}
S \sim  e^{i({\bf k} \cdot {\bf x} - \omega t)} \,\ .
\label{eq:wave0}
\end{equation}
%...................................
Typically such a solution may exist only if $\omega$ and ${\bf k}$ are related by an appropriate equation, called the \emph{dispersion relation}.
If either $ {\bf k}$ or $\omega$ develop imaginary parts leading to positive real arguments in the exponential, the solution  
is expected  to blow in space or time, thus signaling  an ``instability''. Of course, 
the growth is not exponential forever, as it would break the linear approximation. It is, however, 
appropriate to diagnose the onset of instabilities by studying linear stability.  
It should be emphasized that not all modes with complex $ {\bf k}$ or $\omega$ correspond to an instability: the appearance of a positive real term in the exponential argument of Eq.\,(\ref{eq:wave0}) would be a naive and not sufficient criterion. The purpose of
this paper is to discuss more refined criteria to assess the onset of instabilities and to classify them, building upon
the dispersion relation approach proposed in \cite{Izaguirre:2016gsx} for SN neutrinos, and on formally analogous methods developed for unstable phenomena in other contexts, such as plasma physics and fluid dynamics \cite{landau}. 
The conditions for fast flavor conversions will be elucidated by means of both analytical calculations and numerical simulations.

The outline of our paper is as follows. In Sec.\,2 we present the equations of motion of the dense neutrino gas in their complete but linearized form.  In Sec.\,3 we discuss the dispersion relation for the neutrino ensemble and present a kinematical
classification of the possible instabilities of the system. In Sec.\,4 we present the  theory of the instabilities, based on the asymptotic properties
of the Green's function of the problem. In Sec.\,5 we apply this general theory
to a simplified two-beam model. We also  interpret the possible dispersion relations
 in terms of ``particle-like'' vs ``tachyon-like' behaviors. We compare our analytical predictions   with numerical solutions of the linearized equations of motion.  
Finally in Sec.\,6 we summarize  our results and discuss  
the potential consequences for flavor conversions in SNe and neutron star mergers.

%%%%%%%%%%%%%%%%%%%%%%%%%%%%%%%%%%%%%%
\section{Flavor evolution and dispersion equation at high neutrino density}
%%%%%%%%%%%%%%%%%%%%%%%%%%%%%%%%%%%%%%%%%%%%%%%%%

In the absence  of external forces and collisions, 
the dynamics
of the  space-dependent $\nu$ occupation numbers or Wigner function $\varrho_{{\bf p}, {\bf x},t}$
with momentum ${\bf p}$ at position ${\bf x}$ and time $t$ is ruled by the following equations of motion (EoMs)\,\cite{Sigl:1992fn,Strack:2005ux} 
%..........................................................
\begin{equation}
\partial_t \varrho_{{\bf p}, {\bf x},t} + {\bf v}_{\bf p} \cdot \nabla_{\bf x} \varrho_{{\bf p}, {\bf x},t} 
= - i [\Omega_{{\bf p}, {\bf x},t}, \varrho_{{\bf p}, {\bf x},t}] 
\,\ ,
\label{eq:eom}
\end{equation}
%........................................................
where, in the Liouville operator on the left-hand side,  the first term accounts for explicit time
dependence, while the second term, proportional to the neutrino velocity ${\bf v_p}$, encodes the spatial dependence due to particle free streaming. In the absence of oscillations, the right-hand-side would be zero and the EoM would reduce to a Vlasov continuity equation. In the presence of oscillations, the matrix $\Omega_{{\bf p}}$ is the Hamiltonian
%...............................................................
\begin{equation}
\Omega_{{\bf p}}= \Omega_{{\rm vac}} + \Omega_{\rm MSW} + \Omega_{\nu\nu} \,\ ,
\label{eq:ham}
\end{equation}
%............................................................
containing the vacuum, matter and self-interaction terms, that leads to the evolution of $\varrho_{{\bf p}}$ over space and time. Here and in the following, to lighten our notation, we drop the subscripts ${\bf x}$ and $t$.

For our purposes, Eq.\,(\ref{eq:ham}) can be simplified in an effective two-generation scenario, since only two flavors $(\nu_e,\,\nu_x)$ are involved and, moreover, the relevant dynamics near the supernova core will be shown to be independent on the vacuum oscillation parameters which just trigger (but do not govern) fast conversions. The vacuum term is thus the matrix
$\Omega_{{\rm vac}}= \textrm{  diag}(-\Delta m^2 /4E, +\Delta m^2 /4E)$ 
in the mass basis, where $E=|{\bf p}|$ for 
ultra-relativistic neutrinos. 
For antineutrinos, the EoMs are the same but with the replacement $\Omega_{{\rm vac}} \to - \Omega_{{\rm vac}}$, thus it is convenient to think of antineutrinos of energy $E$ as neutrinos of energy $-E$, making their EoMs identical. The matter term in Eq.\,(\ref{eq:ham}) reads
%...................................................................................
\begin{equation}
\Omega_{\rm MSW}=  \lambda\,\textrm{  diag} (1,0) \,\ ,
\end{equation}
%......................................................................
in the weak interaction basis, where $\lambda =\sqrt{2} G_F n_e$. 
Finally, the term due to $\nu-\nu$ interactions is given by
%........................................................
\begin{equation}
\Omega_{\nu\nu} = \sqrt{2} G_F \int \frac{d^3 {\bf q}}{(2 \pi)^3} ({\varrho_{\bf q}} - {\bar\varrho_{\bf q}}) (1 -{\bf v}_{\bf p}\cdot {\bf v}_{\bf q}) \,\ ,
\end{equation}
%..........................................................
where the angular factor $(1 -{\bf v}_{\bf p}\cdot {\bf v}_{\bf q})$ leads to \emph{multi-angle} effects\,\cite{Duan:2006an}, as neutrinos moving on different trajectories experience different self-interaction potentials.
Concerning the density matrix $\varrho_{\bf p}$ in two-flavor case we can write it in the weak-interaction basis as
%...........................................................
\begin{equation}
\varrho = \frac{f_{\nu_e} + f_{\nu_x}}{2}  
\left(\begin{array}{cc} 1 & 0 \\
0 & 1 \end{array} \right)
+  \frac{f_{\nu_e} - f_{\nu_x}}{2} 
\left(\begin{array}{cc} s & S \\
S^* & -s \end{array} \right) \,\ ,
\end{equation}
%.......................................................... 
where $f_{\nu_e}$ and $f_{\nu_x}$ are the initial occupation numbers.
The complex scalar field $S_{\bf p} (t, {\bf x})$ represents the $\nu_e\nu_x$ flavor coherence for the mode ${\bf p}$, while the real field
$s_{\bf p} (t, {\bf x})$ satisfies $s_{\bf p}^2  +|S_{\bf p}|^2=1$.  
Note that since we are assuming the ${\bar \nu}$ have negative energy and negative $\varrho$, the 
 ${\bar \nu}$  coefficients are $-(f_{\bar\nu_e} + f_{\bar\nu_x})/{2}$ 
 and $-(f_{\bar\nu_e} - f_{\bar\nu_x})/{2}$.
 Neutrinos are produced as flavor eigenstates and no flavor mixing occurs as long as $S_{\bf p}=0$. 

While the self-induced flavor evolution described by Eq.\,(\ref{eq:eom}) is a nonlinear problem,
the onset of these conversions can be examined by linearizing the equations, observing that  
$|S| \ll 1$ initially and that $s = \sqrt{1 -|S|^2}\simeq1$ to linear order in $S$. 
In the
context of fast conversions (where $\mu\gg\omega_{\rm vac}$ dominates), the $\Omega_{{\rm vac}}$ term just produces the initial seed $|S|\neq   0$ and then becomes irrelevant, so that we can take 
$\Omega_{{\rm vac}} =0$ hereafter. In this limit, the energy $E$ disappears from the
EoMs, and the modes of $S$ can be labeled by the same unit vector ${\bf v}$ (with $|{\bf v}|\approx c=1$) for both $\nu$ and $\bar\nu$.

Assuming that the occupation numbers as well as the matter density are homogeneous and stationary within the test volume, one gets the following linearized equations
%....................................................
\begin{eqnarray}
i(\partial_t + {\bf v} \cdot \nabla_{\bf x}) S_{\bf v} &=& \left[ \lambda + \int d\Gamma^\prime (1-{\bf v}\cdot {\bf v}^\prime)G_{\bf v^\prime}  \right]  S_{\bf v}\nonumber \\
&-&\int d\Gamma^\prime (1-{\bf v}\cdot {\bf v}^\prime)G_{\bf v^\prime}  S_{\bf v^\prime} \,\ ,
\label{linear}
\end{eqnarray}
%..............................................................
where $d\Gamma= d {\bf v}/4 \pi$ and $G_{\bf v}$ is the angle distribution of the electron lepton number (ELN) carried by neutrinos\,\cite{Izaguirre:2016gsx}, namely
%............................................................................
\begin{equation}
G_{\bf v} = \sqrt{2} G_F \int_{0}^{\infty}\frac{dE\,E^2}{2 \pi^2}\left[f_{\nu_e}(E,{\bf v})-f_{\bar\nu_e}(E,{\bf v}) \right] \,\ ,
\label{eq:eln}
\end{equation}
%..............................................................................
where the $f$'s represent the neutrino distributions in energy and in emission angle.
{The corresponding ELN potential in Eq.\,(\ref{linear}) is given by 
$\mu= \int d\Gamma G_{\bf v}$, with an associated current 
${\mathbf j}= \int d\Gamma  G_{\bf v}  {\bf v}$. 
For typical SN neutrino densities\,\cite{Chakraborty:2011gd} one numerically finds
%............................
\begin{equation}
\mu = \sqrt{2}G_F (n_{\nu_e}-n_{{\bar\nu}_e})  \simeq
6 \,\textrm{m}^{-1}\frac{(n_{\nu_e}-n_{{\bar\nu}_e}) }{10^{31}\,\ \textrm{cm}^{-3}} \,\ ,
\label{eq:dens}
\end{equation}
%..............................
corresponding to a time scale of $ {\mathcal O}(1)$ ns.
}
Note that in the previous equation we are assuming $f_{\nu_x}(E,{\bf v})=f_{\bar\nu_x}(E,{\bf v})$, otherwise one should add a term
$-[f_{\nu_x}(E,{\bf v})-f_{\bar\nu_x}(E,{\bf v})]$.

A further simplification occurs in studying flavor conversions at distances from the SN neutrinosphere much smaller than 
its radius of ${\cal{O}}(10)\,{\rm km}$, since the curvature of the emitting surface becomes irrelevant. 
Then one can simply use Cartesian components for the velocity,
\begin{equation}
{\bf v}=\Big(\sqrt{1-v_z^2}\cos\varphi,\sqrt{1-v_z^2}\sin\varphi, v_z\Big)\,\ ,
\label{eq:veloc}
\end{equation}
where $v_z=\cos\vartheta$ is the component along the $z$-axis, and $\vartheta$ and $\varphi$ the zenith and azimuthal angles, respectively. Note that $v_z$ can take negative values, i.e., the zenith angle $\vartheta$ can take values from $0$ up to $\pi$, 
representing neutrinos with trajectories that range from radially outwards to radially inwards, and
not merely up to $\pi/2$ as usually taken in the ``bulb'' model. With the same logic, 
the background matter and neutrino density can be locally assumed as constant. 

\section{Dispersion relation and classification of the instabilities}

\subsection{Dispersion relation}

Starting from the linearized  equations of motion [Eq.\,(\ref{linear})], one can proceed by seeking solutions via the ansatz 
%....................................
\begin{equation}
S_{\bf v}(t,{\bf x}) = Q_{\bf v} e^{i({\bf k} \cdot {\bf x} - \omega t)} \,\  .
\label{eq:wave}
\end{equation}
%...................................
After a global phase transformation
%................................................
\begin{equation}
S_{\bf v}(t,{\bf x}) \to S_{\bf v}(t,{\bf x}) \exp\left[-i t\left(\lambda + \int d\Gamma^\prime G_{\bf v^\prime} \right) + i  \int d\Gamma^\prime {\bf x}\cdot {\bf v}^\prime G_{\bf v^\prime} \right] \,\ ,
\label{eq:gauge}
\end{equation}
%........................................
to remove the matter and neutrino density terms in the square-bracketed term on the r.h.s.\ of  Eq.\,(\ref{linear}) (see\,\cite{Izaguirre:2016gsx}), one obtains 
%.................................
\begin{equation}
(\omega-{\bf k} \cdot {\bf v})Q_{\bf v}= -\int d\Gamma ^\prime (1-{\bf v}\cdot {\bf v}^\prime)G_{\bf v^\prime}  Q_{\bf v^\prime} \,\ .
\end{equation}
%..........................................
While this transformation makes the equation look simpler, one must be careful that the modes labelled by ${\bf k}=0$ or $\omega=0$ after the transformation do not correspond to what were homogeneous or stationary modes.

The above equation has solutions only if 
\begin{equation}
\label{matrixeq}
\renewcommand*{\arraystretch}{1.8}
\setlength{\delimitershortfall}{0pt}
{\rm det}\begin{bmatrix}
1+\int d\Gamma \frac{G_{\bf v}}{\omega-\bf{v}\cdot\bf{k}} &\int d\Gamma \frac{v_xG_{\bf v}}{\omega-\bf{v}\cdot\bf{k}}&\int d\Gamma \frac{v_yG_{\bf v}}{\omega-\bf{v}\cdot\bf{k}}&\int d\Gamma \frac{v_zG_{\bf v}}{\omega-\bf{v}\cdot\bf{k}}\\
\int d\Gamma \frac{v_x G_{\bf v}}{\omega-\bf{v}\cdot\bf{k}}&-1+\int d\Gamma \frac{v_x^2G_{\bf v}}{\omega-\bf{v}\cdot\bf{k}}&\int d\Gamma \frac{v_x v_y G_{\bf v}}{\omega-\bf{v}\cdot\bf{k}}&\int d\Gamma \frac{v_x v_z G_{\bf v}}{\omega-\bf{v}\cdot\bf{k}}\\
\int d\Gamma \frac{v_y G_{\bf v}}{\omega-\bf{v}\cdot\bf{k}}&\int d\Gamma \frac{v_x v_y G_{\bf v}}{\omega-\bf{v}\cdot\bf{k}}&-1+\int d\Gamma \frac{v_y^2G_{\bf v}}{\omega-\bf{v}\cdot\bf{k}}&\int d\Gamma \frac{v_y v_z G_{\bf v}}{\omega-\bf{v}\cdot\bf{k}}\\
\int d\Gamma \frac{v_z G_{\bf v}}{\omega-\bf{v}\cdot\bf{k}}&\int d\Gamma \frac{v_x v_z G_{\bf v}}{\omega-\bf{v}\cdot\bf{k}}&\int d\Gamma \frac{v_y v_z G_{\bf v}}{\omega-\bf{v}\cdot\bf{k}}&-1+\int d\Gamma \frac{v_z^2 G_{\bf v}}{\omega-\bf{v}\cdot\bf{k}}
\end{bmatrix}=0 .
\end{equation}
Note that in\,\cite{Izaguirre:2016gsx} the previous equation has been written as
%.................................................
\begin{equation}
\textrm{det}\left[\Pi^{\mu \nu} (\omega, {\bf k}) \right] =0 \,\ ,
\label{eq:dispmain}
\end{equation}
%.............................................
in terms of a ``polarization tensor''
%...................................................
\begin{equation}
\Pi^{\mu \nu} = \eta^{\mu \nu} +\int \frac{d{\bf v}}{4 \pi}G_{\bf v} \frac{v^{\mu}v^{\nu}}{\omega-{\bf k} \cdot {\bf v}} \,\ ,
\end{equation}
%................................................
with $v^{\mu}=(1,{\bf v})$ and $\eta^{\mu\nu}=\textrm{diag}(+1,-1,-1,-1)$ being the metric tensor.

Thus, wave-like solutions as in Eq.\,(\ref{eq:wave}) exist only if $\omega$ and ${\bf k}$ (``normal modes'') are appropriately related as per Eq.\,(\ref{matrixeq}) [or, equivalently, Eq.\,(\ref{eq:dispmain})], which is called the 
``dispersion relation'' of the system. 
For the sake of simplicity, hereafter we shall assume that the relevant dynamics involves only one space coordinate $z$ and its conjugate  $k$,
\begin{equation}
\mathbf{x} = z\ ,\ \mathbf{k}=k\ ,
\end{equation}
and that instabilities do not grow in $(x,\,y)$ plane orthogonal to the flow along $z$. 
The study of multidimensional instabilities is beyond the scope of this work.

%%%%%%%%%%%%%%%%%%%%%%%%%%%%%%
\subsection{Types of Instabilities}
%%%%%%%%%%%%%%%%%%%%%%%%%%%%%%%%
\label{sec:class}

For a system described by a generic linear differential equations in space and time,
%...........................................
\begin{equation}
D\left(i\frac{\partial}{\partial t},-i\frac{\partial}{\partial z}\right){S}(z,\,t)=0 \,\ .
\label{eq:dispspace}
\end{equation}
%................................................
The ansatz in Eq.\,(\ref{eq:wave}) leads to a dispersion relation of the kind, 
%...................................
\begin{equation}
D(\omega,\, k)=0 \,\ ,
\label{disp}
\end{equation}
%........................................
which is solved by normal modes, providing an expansion basis to decompose any space-time perturbation $S$.

One can learn a great deal about the stability properties of the system, by studying the conditions which (do not) lead
to imaginary parts for $\omega$ or $k$ in the dispersion relation. In general, the solutions can be cast either in the form
%....................................
\begin{equation}
\omega= \Omega({k})\in\mathbb{C}\quad{\rm with}\quad k\in\mathbb{R} \,\ ,
\label{eq:omega}
\end{equation}
%..................................
or in the form
%....................................
\begin{equation}
{k}= {K}(\omega) \in \mathbb{C}\quad{\rm with}\quad \omega\in\mathbb{R} \,\ . 
\label{kappa}
\end{equation}
%..................................

Solutions in terms of Eq.\,(\ref{eq:omega}) involve 
a \emph{temporal}  
stability analysis where one searches for growing modes in time
for a wave propagating with real wavenumber ${k}$. If not only $k$ but also $\Omega({k})$ is real, the wave is stable, while if 
$\Omega({k})$ develops a positive imaginary part, it can become unstable at large time $t$. On the other hand, solutions in terms of Eq.\,(\ref{kappa}) involve 
a \emph{spatial} stability analysis. In this case, if ${K}(\omega)$  develops an imaginary part for some real $\omega$,  exponentially growing or decaying solutions would emerge as a function of the distance $z$.

Following the seminal work in \cite{Sturrock:1958zz}, the stability of a linearized system can be studied in terms of
the long-time behavior of a generic perturbation $S$, expressed either as 
``space-like'' wavepacket {(when the perturbation can grow with time, while propagating in space)}
 %..........................
 \begin{equation}
S(z,t) = \int d{ k} \,g_{{k}} e^{i({k}  {z}-\Omega ({k}) t)}\, ,
\label{eq:time}
\end{equation}
%......................................................
or as a ``time-like'' wavepacket {(when the perturbation can grow {along a spatial direction, as time elapses})},
%......................................................
\begin{equation}
S(z,t) = \int d\omega \,f_{\omega} e^{i({K}(\omega)  {z}-\omega t)}\, ,
\label{eq:space}
\end{equation}
%......................................................
depending on the problem under study, e.g., the response to a space-localized perturbation, or to a forcing harmonic frequency, etc. By means of general arguments about the convergence properties of integrals with oscillating integrands, it was shown
\cite{Sturrock:1958zz} that the perturbation asymptotics generally fall into one of the following four categories, which 
have been almost universally adopted in the  literature on linear instabilities: 
\begin{itemize}
\item[$\bullet$]  \emph{absolute instability}, if
the perturbation blows up ``on site'' and around;
\item[$\bullet$]\emph{convective instability}, if the 
perturbation decays locally, but blows {up} elsewhere as it moves away;
\item[$\bullet$] \emph{complete stability}, if the perturbation is neither enhanced nor damped; 
\item[$\bullet$] \emph{stability with damping}, if the perturbation decays (exponentially) in space.%
%---------
\footnote{
For electromagnetic disturbances, this case would lead to the so-called  ``evanescent waves'' and ``non-transparency'' phenomena \protect\cite{landau}.}
%----------
\end{itemize}

\begin{figure}[!t]
\begin{centering}
\includegraphics[width=0.4\textwidth]{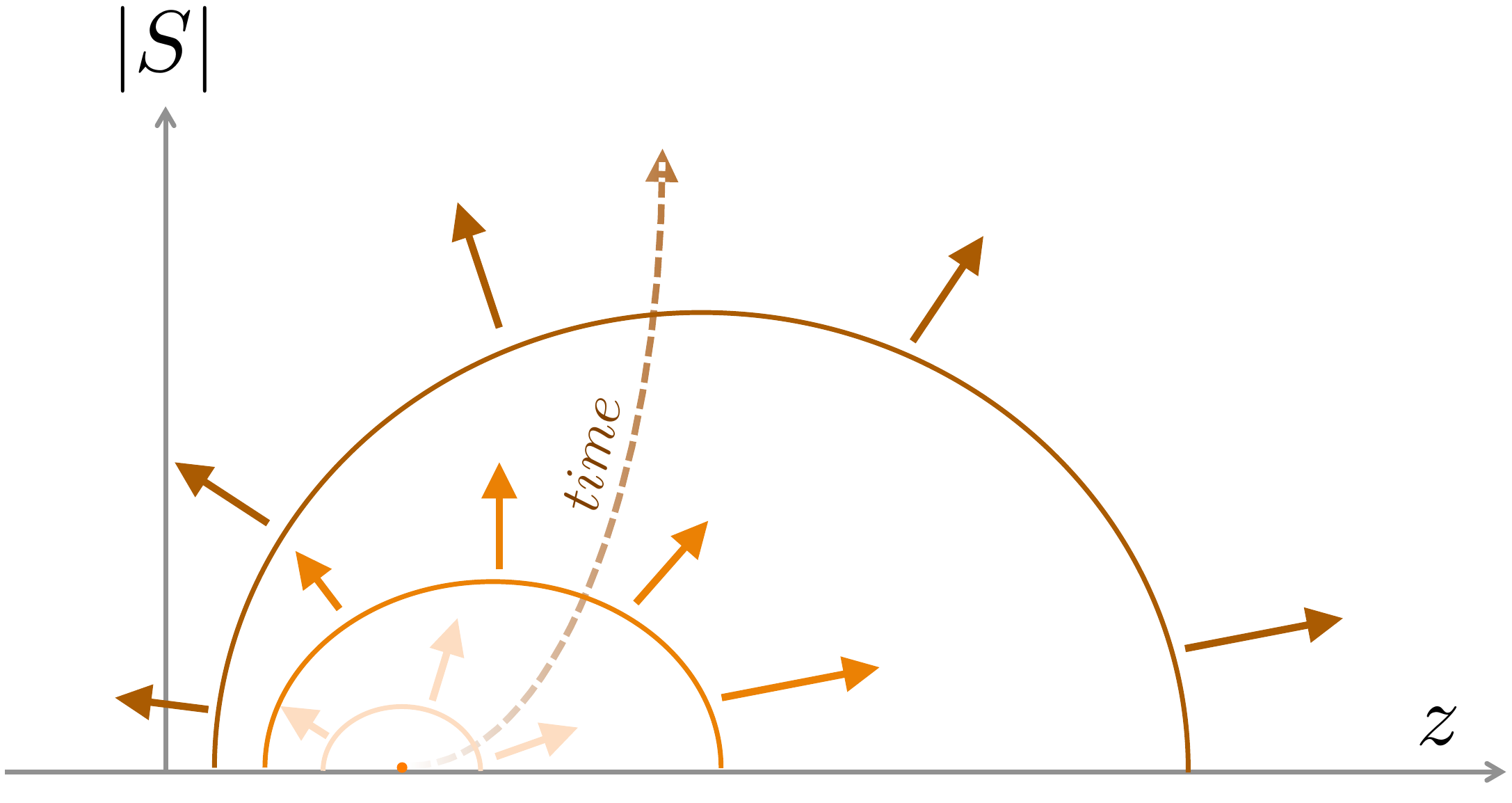}\,\,\,\,\,\,\,\,\,\,\,\includegraphics[width=0.4\textwidth]{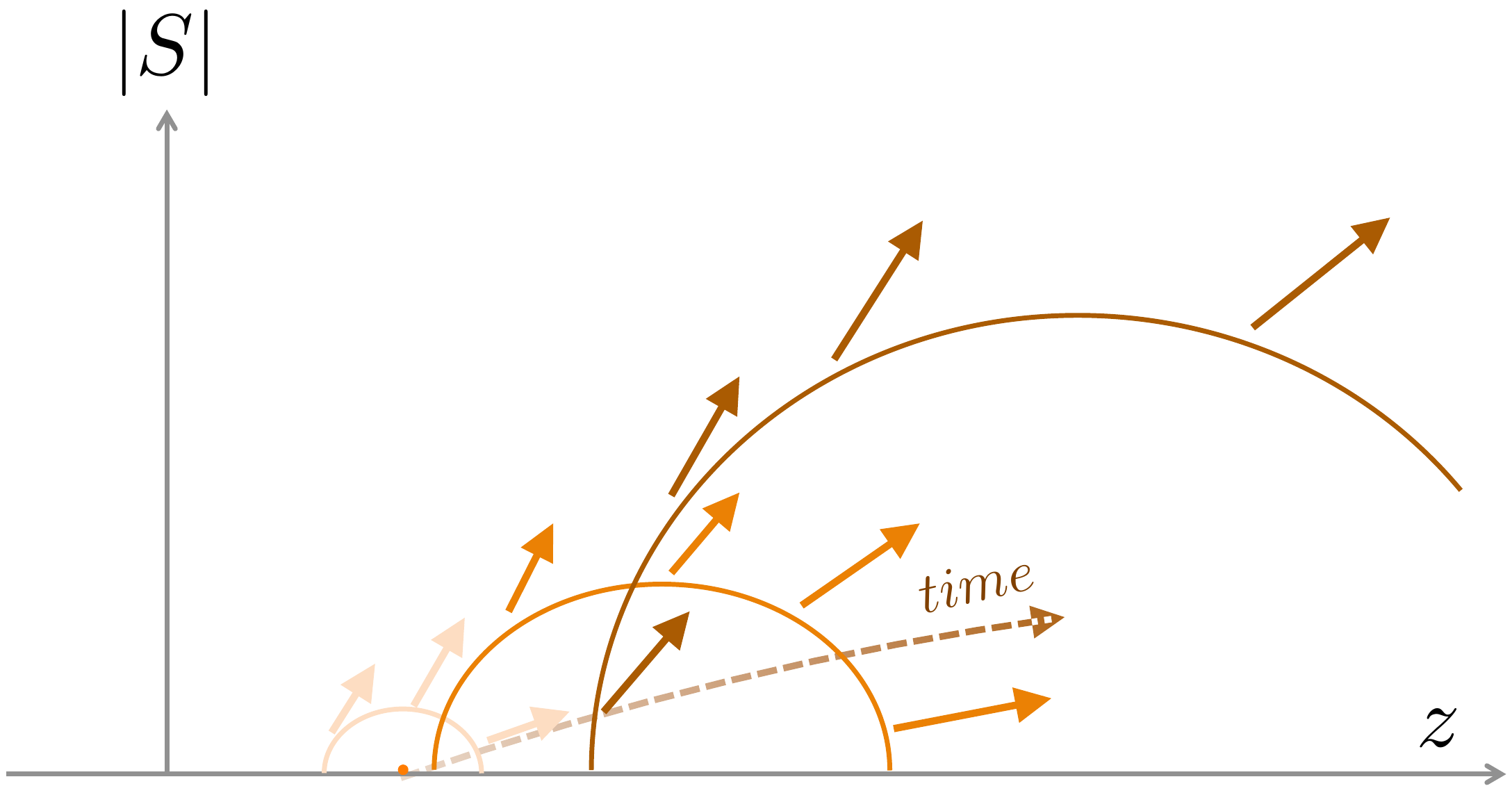}
\end{centering}
\caption{Sketches of {amplitude} of the wave function $S(z,t)$ {at three moments in time (darker colors denote later times)}, for absolute (left panel) and convective (right panel) instability.}
\label{fig:abs-conv}
\end{figure} 

The time-asymptotic behavior of the function $S(z,t)$ for absolute and convective unstable cases is sketched in Fig.\,\ref{fig:abs-conv}. 
In \cite{Sturrock:1958zz} the above (un)stable cases were also set in correspondence with the emergence of complex values of $(\omega,\,k)$
in the dispersion relation: 
\begin{itemize}
\item[$\bullet$] {absolute instability}, when $\omega$ can be complex {for real $k$},  but $k$ remains real {for all real $\omega$};
\item[$\bullet$] {convective instability}, when  $\omega$ can be complex {for real $k$}, and $k$ can be complex {for real $\omega$} ;
\item[$\bullet$] {complete stability}, when  $\omega$ is real {for all real $k$}, and $k$ is real  {for all real $\omega$};  
\item[$\bullet$] {stability with damping}, when $\omega$ is real for all real $k$, but $k$ can be complex for real $\omega$.
\end{itemize}
We shall see an example of such correspondence in the simple two-mode system of Sec.\,V.
However, the above diagnostics in terms of complex or real  $(\omega,\,k)$, despite being useful in simple systems, does not lead 
to quantitative insights and, most importantly, may not be
general enough to capture more complex situations.
{Already from the above classifications of instabilities we realize that the $z$ and $t$ variables
(as well as their conjugates $k$ and $\omega$), 
do not play symmetrical roles in the dispersion relation approach.} 
Indeed, perturbations causally propagate forward in time (and are absent for $t<0$), while they can propagate both downstream ($z>0$) and upstream ($z<0$) in space, involving Laplace and Fourier transforms for 
the conjugate variables $(t,\,\omega)$ and $(z,\,k)$, respectively. Thus, 
while a positive imaginary part of $\omega$ is a clear indicator of a time-growing instability, the sign of the imaginary part of $k$  is not an obvious 
diagnostic of space-growing (or decaying) perturbations. In addition, for impulsive perturbations 
one is usually interested in the local $(z=0)$ behavior of $S$ at large time ($t\to \infty$), while for steady harmonic forcing one focuses on space growth. 
Finally, multi-valued solutions may be more easily tractable in terms of $K(\omega)$ or $\Omega(k)$, depending on the
system. Thus, both temporal and spatial analyses may be needed, in order to fully understand the  instability conditions 
of the system, see e.g., \cite{landau,schmid}. A rather general approach, originally developed in the context
of plasma physics \cite{briggs, bers} and further extended also to
fluid dynamics \cite{landau,schmid}, is described in the following section.

%%%%%%%%%%%%%%%%%%%%%%%%%%%%%%
\section{Theory of Instability}
%%%%%%%%%%%%%%%%%%%%%%%%%%%%%%%%
 \label{sec:briggs}

We discuss below the general theory that leads to criteria distinguishing the different 
types of instabilities presented in the previous section, {the so-called Briggs-Bers criterion} \cite{briggs,bers}. We follow 
the Green's function formulation  given in\,\cite{landau} and expanded in later reviews, 
see e.g.\ \cite{huerremonk,huerre,chen}.

 \subsection{Green's Function}

The Green's function ${G}(z,\,t)$ encodes the response of the linear system governed by
Eq.\,(\ref{eq:dispspace}) to an impulsive forcing (or ``source'') term, 
%......................................
 \begin{equation}
 D\left(i\frac{\partial}{\partial t},-i\frac{\partial}{\partial z}\right) { G}(z,\,t)= \delta(z) \delta(t) \,\ .
\label{eq:green}
 \end{equation}
%....................................................
 If ${G}(z,\,t)$ is known, the response $S(z,\,t)$ to a generic forcing $f(z,\,t)$ 
is obtained by a double convolution of $G$ and $S$. In terms of conjugate variables, the Green's function is simply the inverse of the dispersion function,
%......................................
 \begin{equation}
 D(\omega,\,k) { G}(\omega,\,k)= 1 \,\ ,
\label{eq:greenconj}
 \end{equation}
%....................................................
and the response $S$ is given by $S(\omega,\,k)=G(\omega,\,k)f(\omega,\,k)$. 

\begin{figure}[!t]
\begin{centering}
\includegraphics[width=0.4\textwidth]{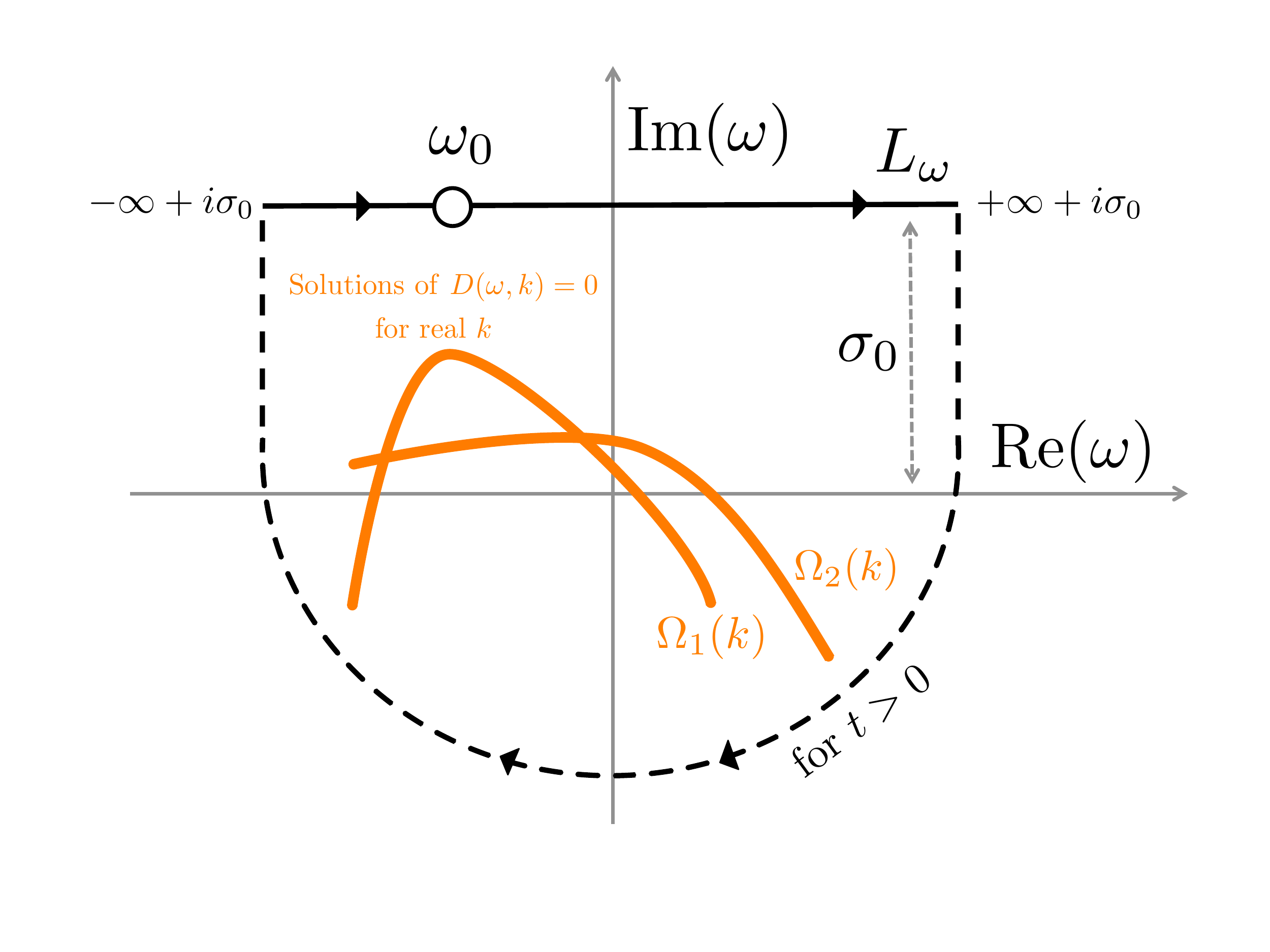}\,\quad\includegraphics[width=0.4\textwidth]{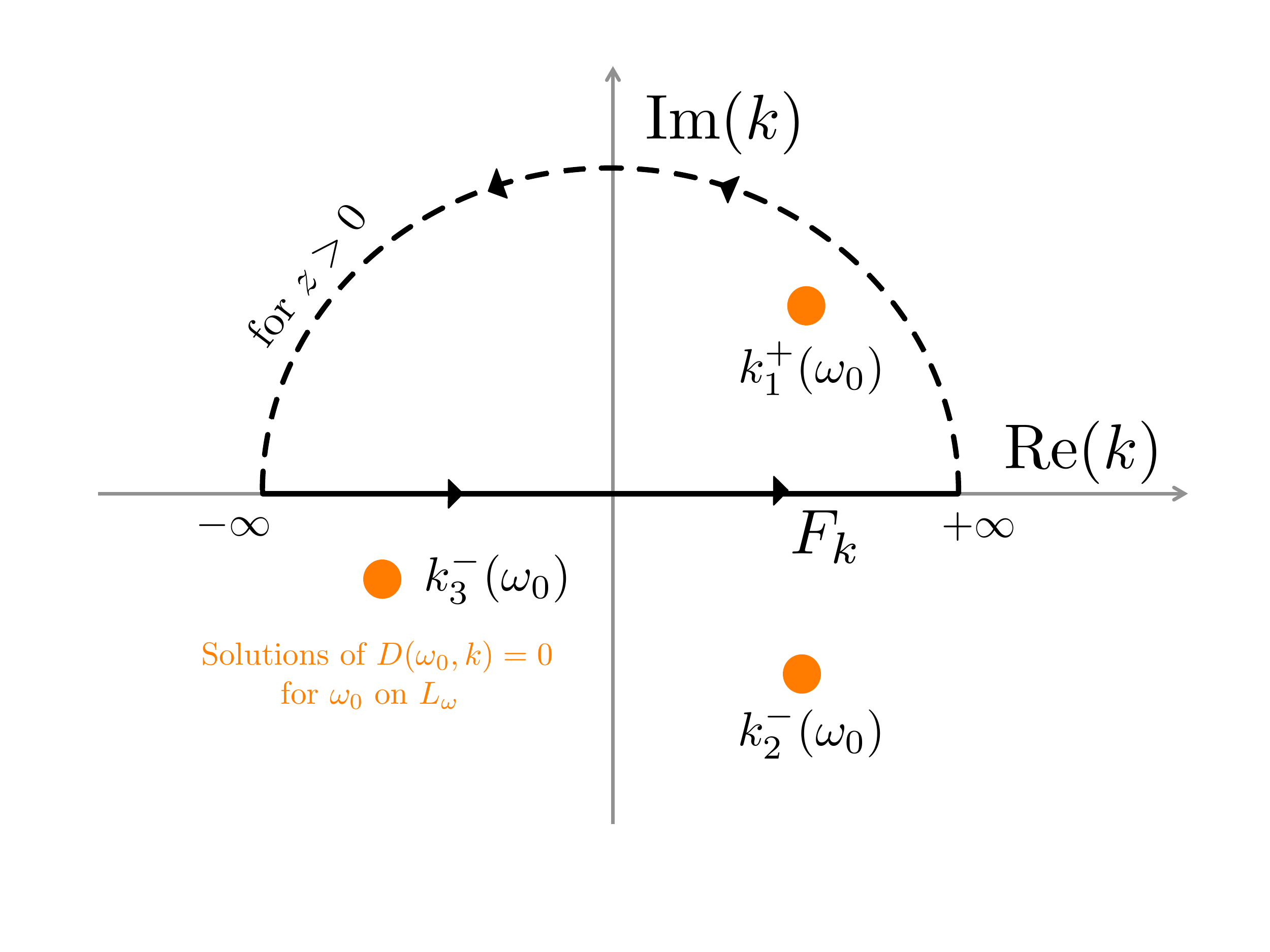}
\end{centering}
\caption{Sketches of the $\omega$-contour  (left panel) 
and $k$-contour (right panel) in Eq.\,(\protect\ref{eq:transform}), together with generic poles. The $\omega$-contour $L_\omega$ is a straight line above all poles of $G(\omega,k)$ for \emph{any} real $k$, shown as the thick curves, and closes clockwise over the lower half-circle for $t>0$ (anticlockwise over upper half-circle for $t<0$). The $k$-contour $F_k$ is along the real $k$-axis, and closes upward for $z>0$ (downward for $z<0$). The dots denotes poles of $G(\omega,k)$ for \emph{given} $\omega_0$ on $L_\omega$. None of these $k$-poles are on the real $k$-axis because of the choice of $L_\omega$.} 
\label{fig:contour}
\end{figure}

According to Eq.\,(\ref{eq:greenconj}), the Green's function $G(z,\,t)$ admits an integral representation as a double (Laplace-Fourier) inverse transform of $1/D$, 
%.........................................
\begin{equation}
{G}(z,\,t) = \frac{1}{(2\pi)^2} \int_{L_\omega} \int_{F_k} {d\omega} {dk}\, 
\frac{e^{i({k}{z}-\omega t)}}{D(\omega,\,k)} \,\
 \,\ ,
 \label{eq:transform}
\end{equation}
%................................................
which is governed by the poles of {$G(\omega,\,k)$, or equivalently by the zeros of $D(\omega,k)$,} in the complex planes of $k$ and $\omega$.   
The Fourier integration domain $F_k$ is the usual real axis $k\in(-\infty,\,+\infty)$, closed by an upper half-circle for $z>0$ (enclosing a family of poles $k^+$) and by a lower half-circle for $z<0$ (enclosing a family of poles $k^-$), 
in order to ensure convergence. {The Laplace contour $L_\omega$ must 
be shifted to 
$\omega\in(-\infty+i\sigma_0,\,+\infty+i\sigma_0)$, with $\sigma_0$ high enough to lie above all poles $\omega(k)$ with $k\in\mathbb{R}$ (this is often referred to as a Bromwich contour).} In this way, $L_\omega$ can be closed by an upper half-circle to satisfy causality 
($G(z,\,t)$=0) for $t<0$, while for $t>0$ it is closed by a lower half-circle. 
Figure\,\ref{fig:contour} shows a sketch of such integration contours in the $\omega$ and $k$ planes. Note that poles may form discrete or  
(possibly multi-valued) continuous sets, depending on the system under study. 
{In particular, we assume that 
 the integral on $F_k$ in Eq.\,(\ref{eq:transform}) is performed first,
followed by the integral on the $L_\omega$.
Then in the figure, the thick curves  in the $\omega$-plane
correspond to all solutions of the relation $D(\omega_j, k)=0$ for any real $k$ and $j=1,\ldots N$ correspond to the $j$th
branch of the solution. For example, if $D(\omega, k)=0$ is quadratic in $\omega$, as it happens in the two-beam example we discuss later, there are two branches for $\omega$ at every real $k$ as shown here.
Meanwhile, the dots in the $k$-plane correspond to solution of the dispersion relation for \emph{a given} $\omega=\omega_0$ on the isocontour of $\textrm{Im}(\omega)$ on $L_\omega$, i.e.,
$D(\omega_0, k_l(\omega_0))=0$, $l=1,\ldots M$ and $\omega_0$ on $L_\omega$. For example, if $D(\omega, k)=0$ is cubic in $k$ there are three roots for a given choice of $\omega_0$, as shown.}

%%%%%%%%%%%%%%%%%%%%%%%%

General features of $G(z,\,t)$ emerge already in the simple case of a single pole $\omega=\Omega(k)$ for real $k$. For $t>0$, by closing the $L_\omega$ contour  with a lower half-circle (clockwise oriented), the residue at $\Omega(k)$ yields\,\cite{huerremonk,huerre}:
%................
\begin{equation}
G(z,\,t) = \frac{-i}{2\pi}\int_{F_k} dk\; \frac{e^{i(kz-\Omega(k)t)}}{\left.\frac{\partial D}{\partial \omega}\right|_k}\ \,,
\end{equation}
%..............
{with the overall minus sign due to the contour going clockwise.}
The behavior of this integral at $t\to\infty$ along a propagation ray with finite $z/t$,
%..............
\begin{equation}
\label{rayV}
V = \frac{z}{t} = \mathrm{const}
\end{equation}
%..............
can be obtained by the steepest descent method around the point where the phase $(kV-\Omega)t$ is stationary. {Namely the point with coordinates $\hat \omega = \Omega(\hat k)$ and $\hat k$ where}
%.................
\begin{equation}
\label{phaseV}
V = \left.\frac{\partial \Omega}{\partial k}\right|_{\hat{k}}\ .
\end{equation}
%....................
A second-order expansion of the integrand around this stationary point yields the time-asymptotic result (derivation omitted)\,\cite{huerremonk,huerre}:
%...............
\begin{equation}
G(x,\,t) \sim -\frac{e^{i\frac{\pi}{4}}}{(2\pi)^\frac{1}{2}}\frac{e^{i(\hat k z - \Omega(\hat k))t}}{\left.\frac{\partial D}{\partial \omega}\right|_{\hat{k}} \left( \left.\frac{\partial^2 \Omega}{\partial k^2}\right|_{\hat{k}} \cdot t \right)^\frac{1}{2}}\ ,
\end{equation}
%................
which corresponds to a wavepacket with frequencies and wavenumbers related by the ``group velocity'' in Eq.\,(\ref{phaseV}). 

The shape of the wave packet changes with the ray $V$ and, in particular, the phase $(\hat kV-\hat \omega)t$ may acquire an imaginary part $\sigma$ for some $V$,
%-------
\begin{equation}
\sigma(V) = \textrm{Im}(\hat \omega - V\hat k)\ \neq 0
\label{eq:sigma}
\end{equation}
%----- 
If $\sigma(V)\leq 0$  for any $V$, then the flow is  
stable. Conversely, if $\sigma(V)>0$ for some $V$, then the wavepacket amplitude grows as $e^{\sigma t}$.  In particular, if $\sigma >0$ for $V=0$, then the amplitude grows exponentially at $z=0$, i.e., locally (case of absolute instability), 
otherwise it moves away faster than it spreads out (case of convective instability). 
   
One can thus translate the classification of the instabilities discussed in Sec.\,\ref{sec:class} in terms of  
properties of the Green's function along propagation rays $V=z/t$.  In particular, a stable flow implies that
%................................
\begin{align}
\lim_{t\to\infty} {G}(z,t)=0 
\,\ \,\ \,\ \,\ \textrm{  along all rays}  \,\ \,\ \,\ 
V=z/t \,\ ,
\end{align}
%.....................................................
while an unstable flow arises when the wave packet grows along at least for some $z/t$, 
%................................
\begin{equation}
\lim_{t\to\infty} {G}(z,t)=\infty \,\ \,\ \,\ \textrm{  along some  rays}  \,\ \,\ \,\ V=z/t \,\ .
\end{equation}
%.....................................................
Convective or absolute instability can be further distinguished by the local response at $z=0$, i.e., along the ray $z/t=0$.
An unstable flow displays a convective instability if
%................................
\begin{equation}
\lim_{t\to\infty} {G}(z,t)=0 \,\ \,\ \,\ \,\ \textrm{  along the ray}  \,\ \,\ \,\ V=0 \,\ ,
\end{equation}
%.....................................................
otherwise the instability is absolute,
%................................
\begin{equation}
\lim_{t\to\infty} {G}(z,t)=\infty \,\ \,\ \,\ \,\ \textrm{  along the ray}  \,\ \,\ \,\ V=0 \,\ .
\end{equation}
%.....................................................

Note that, in a temporal stability analysis with $\omega=\Omega(k)$ for real $k$, it is simply $\sigma = \textrm{Im}(\Omega(k))$, and the above criteria
would seem to be related only to the position of poles in the complex $\omega$ plane. However, this is often but not always the case, and
a more powerful (geometric) criterion can be envisaged to characterize the instabilities.

\subsection{Geometric Criterion for Instability}

We consider first the time asymptotics of an initially localized disturbance $\delta(z)\delta(t)$,
in order to refine the characterization of absolute and convective instabilities. Then we consider 
the evolution of a localized harmonic forcing $\delta(z)e^{i\omega_0t}$, which helps to refine the
discrimination between convective (unstable) and damped (stable) cases.

\subsubsection{Absolute vs Convective Instability}

\begin{figure}[!t]
\begin{centering}
\,\,\includegraphics[width=0.4\textwidth]{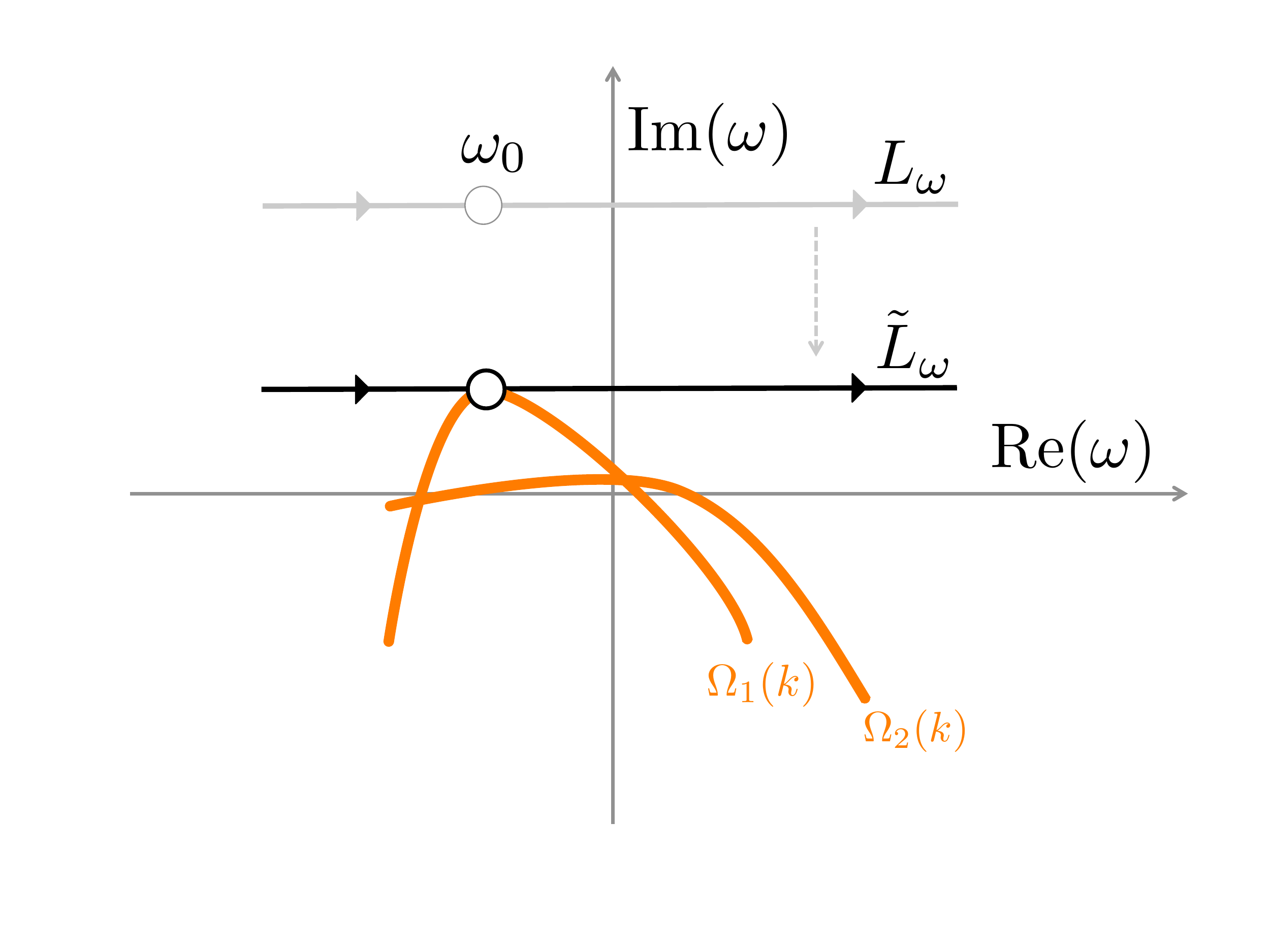}\,\includegraphics[width=0.42\textwidth]{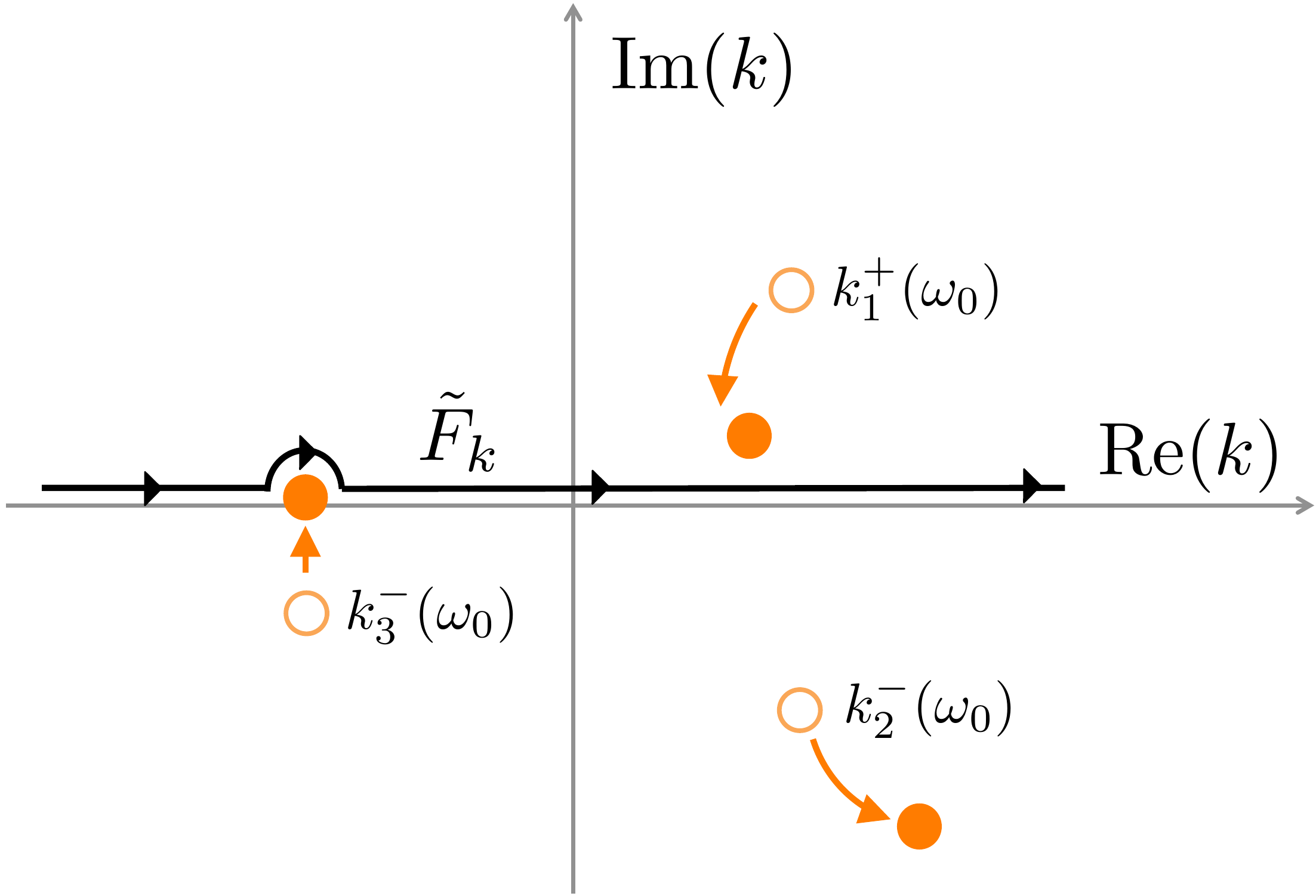}
\end{centering}
\caption{Deformed contours for the analytic continuations of integration in $\omega$ plane (left panel) and $k$-plane (right panel). {The $L_\omega$ contour is lowered towards the real $k$-axis to ascertain the asymptotic behavior of $S$. In doing so, at some point $L_\omega$ must hit a pole if there are $\omega$-poles with $\textrm{Im}(\omega)>0$. When that happens, the $F_k$ contour also encounters a pole and in order to avoid it must be deformed to $\tilde{F}_k$}, as shown in the right panel. {Two possibilities can arise for $\tilde{F}_k$, as shown in the next figure.}
}
\label{fig:contourdef}
\end{figure}

\begin{figure}[!t]
\begin{centering}
\,\,\includegraphics[width=0.26\textwidth]{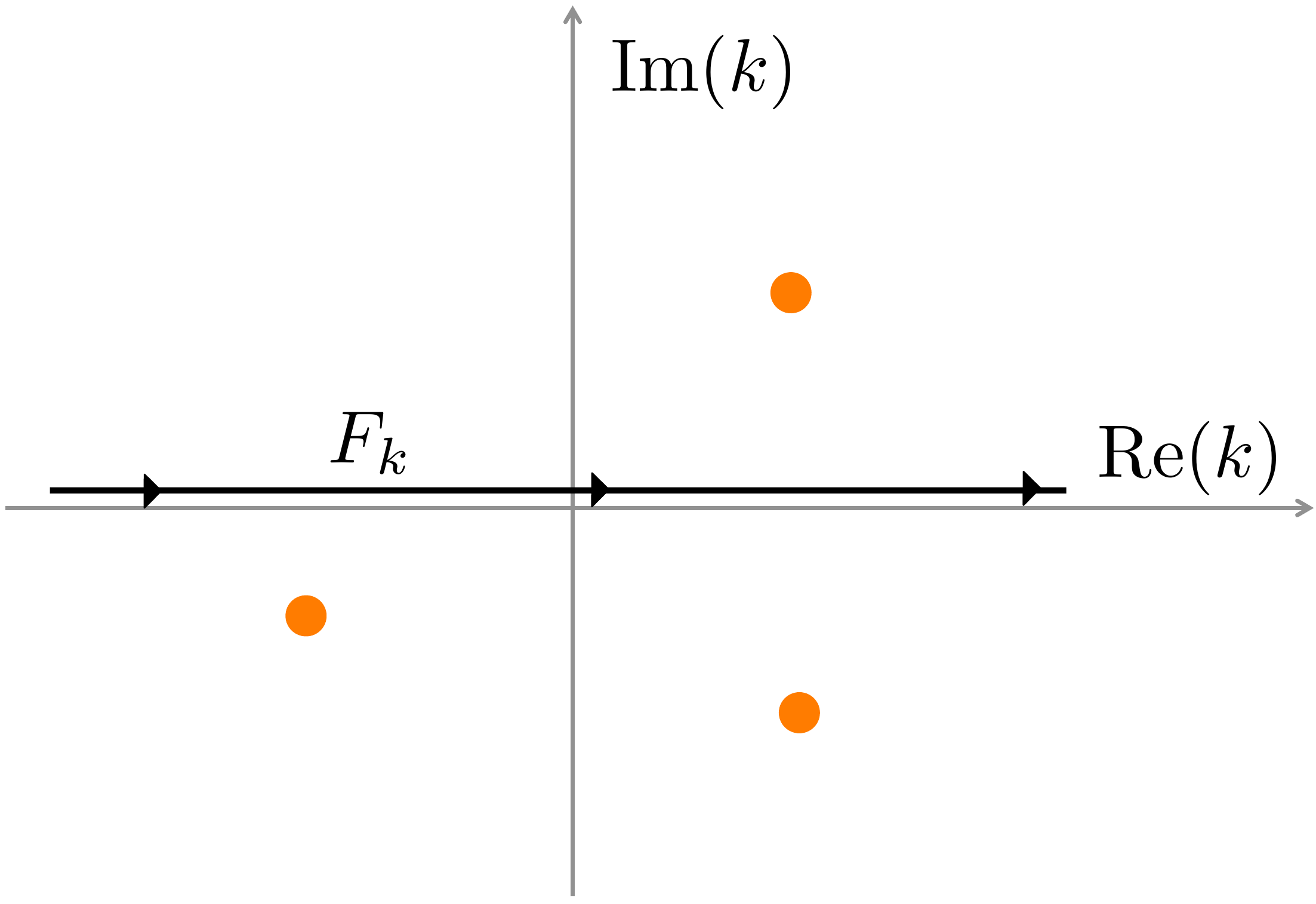}\,\,\,\,\includegraphics[width=0.26\textwidth]{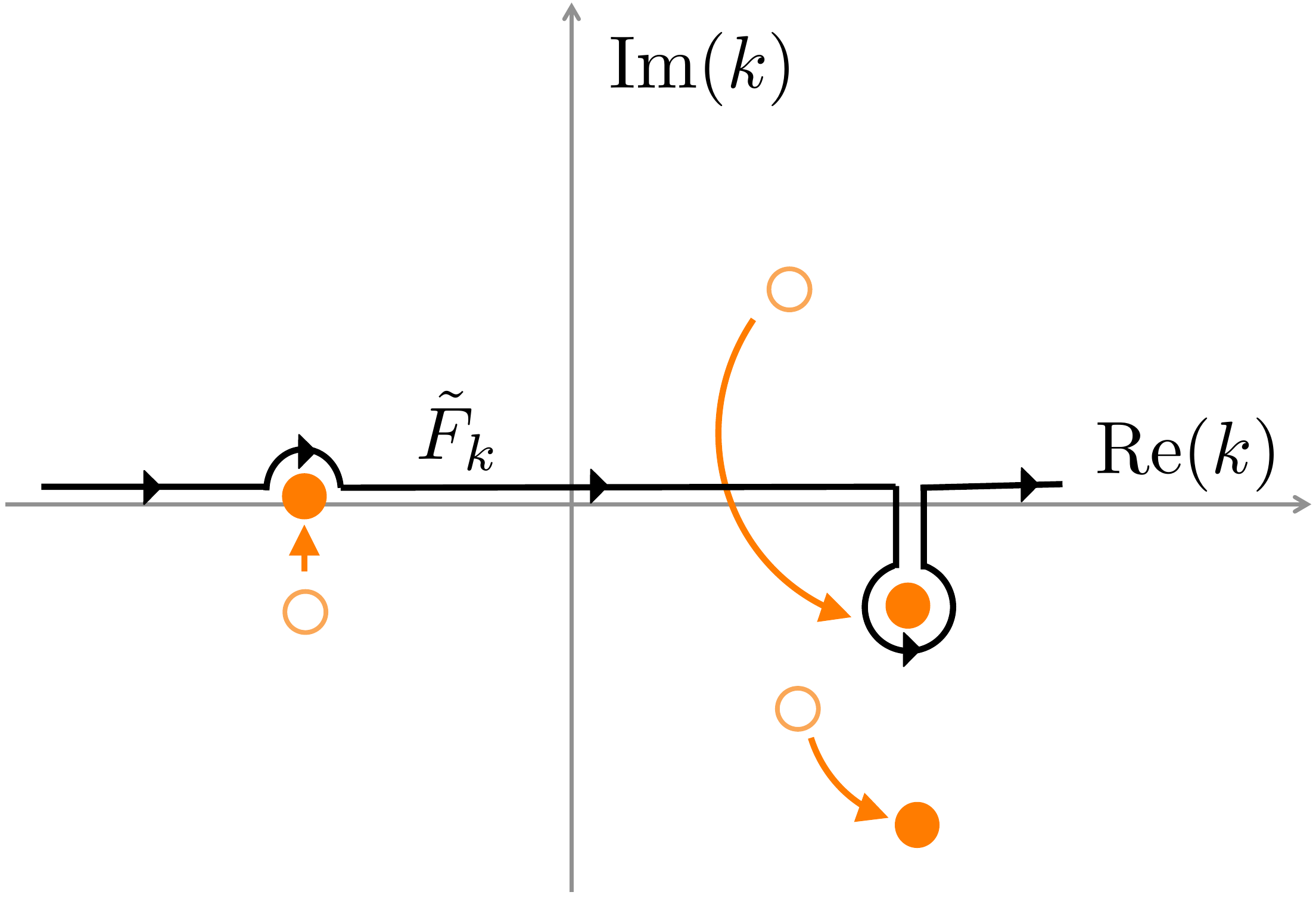}\,\,\,\,\includegraphics[width=0.26\textwidth]{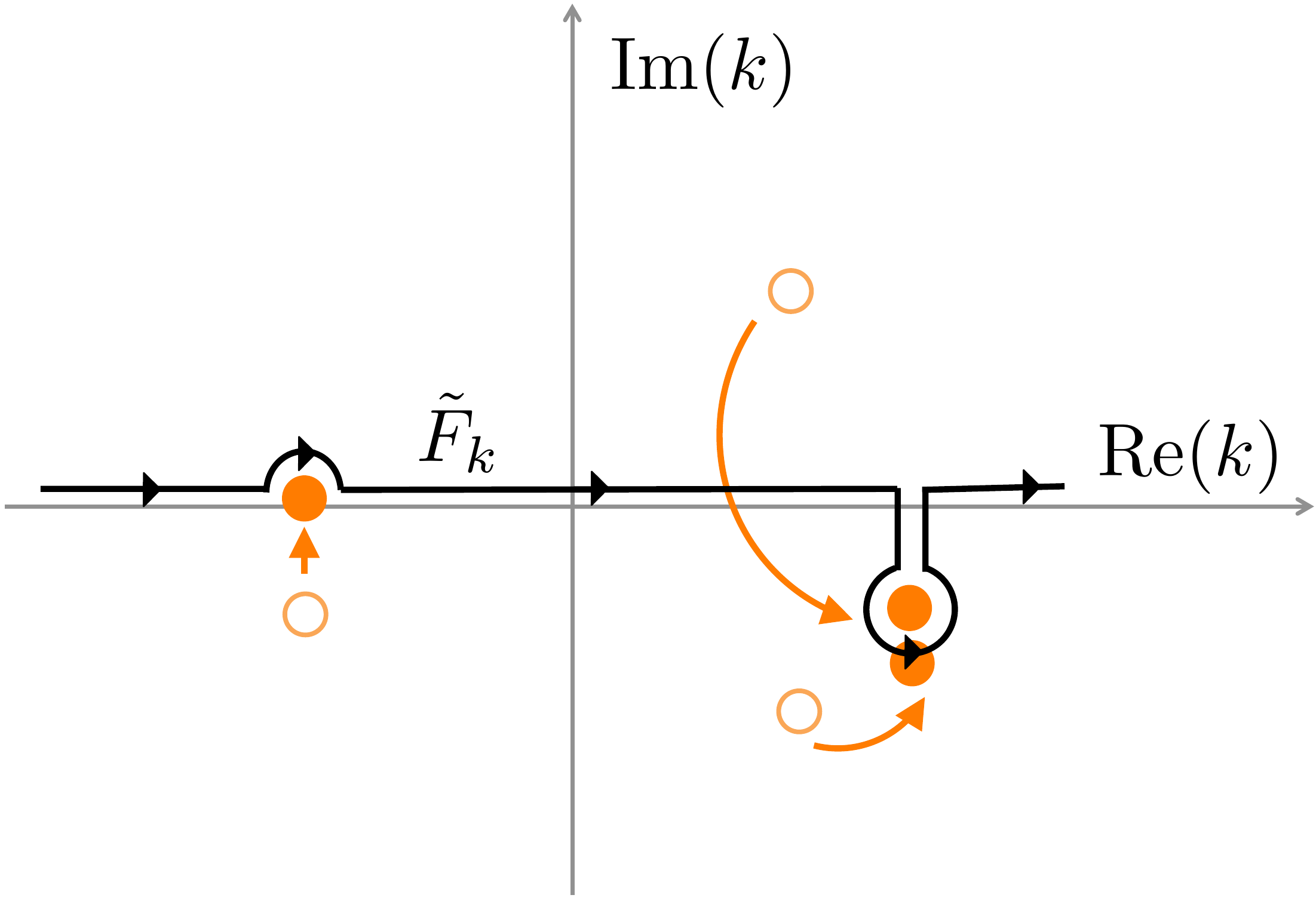}
\end{centering}
\caption{{Roots of the dispersion relation in the $k$-plane for an $\omega_0$ on the original $L_\omega$ contour, e.g., at $\textrm{Im}(\omega) \to \infty$ (left panel). As $L_\omega$ is lowered towards ${\rm Im}(\omega) \to 0$, some pole(s) in the $k$-plane may cross the real $k$-axis but they can be avoided by deforming the  ${F}_k$ contour
 (middle panel). However, some poles pinch the contour $\tilde{F}_k$  and there is no way of avoiding them (right panel). 
 }
}
\label{fig:fincont}
\end{figure}

The criterion is based on two key observations. The first is that the Fourier and Laplace contours in Eq.\,(\ref{eq:transform}) and in Fig.\,\ref{fig:contour} 
can be continuously deformed in the corresponding complex planes, 
as far as they do not cross any pole. Studying the large-time behavior of $G$ implies lowering the $L_\omega$ contour 
towards the half-plane with negative $\textrm{Im}(\omega)$, the integral being then dominated by the highest pole in the complex $\omega$ plane.
{Referring to Fig.\,\ref{fig:contourdef}  we take the contour $L_\omega$ and lower it downward (left panel). If we focus our attention on a specific point on $L_\omega$, denoted by $\omega_0$, we see that as $L_\omega$ is lowered that point gets closer to a pole.
Since the corresponding poles in the $k$-plane, i.e., $k_l=k_l(\omega_0)$, depend on $\omega_0$, they will also move around in the  $k$-plane as $L_\omega$ is moved downward (right panel).
In particular, when the point on $L_\omega$ that was $\omega_0$ eventually hits a pole on the highest branch $\Omega_j$, some pole(s) in $k$-plane must also hit the real $k$-axis. This is because $\Omega_j$ is defined by
$D(\Omega_j, k_l)=0$ with $k_l$ being real. Once $k_l$ becomes real, the Green's function $G(\omega,z)$ is no longer analytical because $k_l$  lies on 
the $F_k$ contour of Eq.\,(\ref{eq:transform}) which is the real $k$-axis. In order to analytically continue $G(\omega,z)$ below the largest value of ${\rm Im}(\Omega_j)$  it is then necessary
to deform the $k$-integration contour around those poles which approach or cross the real $k$-axis, i.e., choose a new contour ${\tilde F}_k$ as shown in 
Fig.\,\ref{fig:contourdef} (right panel).} 
Figure \,\ref{fig:fincont} shows the different possibilities regarding deformation in the complex $k$-plane. 
{{red}{In the left panel we show the roots of the dispersion relation in the $k$-plane for a point $\omega_0$ on the original $L_{\omega}$ contour, i.e., at 
$\textrm{Im}(\omega) \to \infty$.
As $L_\omega$ is lowered towards ${\rm Im}(\omega) \to 0$,  the poles move in the $k$-plane. Some of them can cross the  real $k$-axis but they can be avoided by deforming
the ${F}_k$ contour (middle panel).  However, a singularity occurs if two moving poles come close together and pinch the contour of integration $\tilde{F}_{k}$. This latter
becomes ``stuck'' and cannot be deformed around the two merging poles, signalling an instability. }
}
%If none of the $k$-poles change their upper/lower half-plane the contour $F_k$ doesn't need to be changed and there is no instability (left panel). If some of the $k$-poles poles cross the real $k$-axis before $L_\omega$ can be lowered to the real $\omega$-axis but no poles pinch together, there is a convective instability (middle panel). If, however, two poles pinch the contour $F_k$ before $L_\omega$ reaches the real $\omega$-axis there is an absolute instability (right panel).

%%%%%%%%%%%%%%%%%%%%%%%%
%%%%%%%%%%%%%%%%%%%%%%%

%%%%%%%%%%%%%%%%%%%%%%%%%%%
%%%%%%%%%%%%%%%%%%%%%%%%%

The second observation is that the modes at zero group velocity $V=0$, obtained for some ($\omega_c,\,k_c$), are double roots
of the dispersion relation, since they obey not only $D=0$ but also $\partial D/\partial k =0$ via Eq.\,(\ref{phaseV}). 
The dispersion relation can
be locally expanded as  
%...............................................................
\begin{equation}
D(\omega,\,k) \approx (\omega-\omega_c)\left(\frac{\partial D}{\partial \omega}\right)_c +
\frac{1}{2} (k-k_c)^2\left(\frac{\partial ^2 D}{\partial k^2}\right)_c = 0 \,\ ,
\end{equation}
%..............................................................
which implies that $k=k(\omega)$ has two local solutions $k^\pm$ and that
the pinching point is just $k_c$, where the two poles $k^\pm$ coalesce. The corresponding $\omega$-pole is located at $\omega_c$, with
$\textrm{Im}(\omega_c)$ determining the nature of the instability. 
If $\textrm{Im}(\omega_c)<0$, the contour $L_\omega$ can be safely lowered below the
real axis, so that $G\to 0$ at large time  and the instability is convective. If instead $\textrm{Im}(\omega_c)>0$, the instability necessarily grows in time 
and is absolute. In a nutshell, absolute instabilities arise when modes with zero group velocity develop a positive imaginary part $\textrm{Im}(\omega_c)$, in
which case both downstream and upstream flows ($k^\pm$) mix and blow up. 

The ``pinching'' criterion is consistent with the classification of the previous subsection, but is also more general. In fact, there may be
cases where a double root of $D=0$ is found, but the $k$-poles approach the deformed $F_k$ contour from the same 
side rather than from opposite sides, 
in which case  instabilities do not arise. The criterion is also amenable to numerical and graphical implementations which help to 
identify instabilities:  A cartesian grid spanning the complex $\omega$-plane near $\omega=\omega_c$ is mapped into a deformed grid in the $k$-plane with a saddle point  
at $k=k_c$; vice versa, a cartesian grid in the complex $k$-plane 
around  $k_c$ is mapped into a deformed grid in the $\omega$-plane with a cusp point at $\omega_c$ \cite{schmid}.

\subsubsection{Convectively unstable and Damped stable solutions}

The previous criterion for instabilities was mainly based on a stability analysis in the time domain. 
Further insights can be gained by looking at the evolution of disturbances in space. In particular, 
the damped stable solution introduced in Sec.\,III\,B can be better characterized and separated from the convective unstable solution 
(both decaying away at $z=0$), by 
considering the upstream and downstream
response of the system to a harmonic forcing with real frequency $\omega_f$ and localized at $z=0$, namely,
$f(z,t) \sim \delta(z)e^{-i\omega_f t}$ for $t>0$. In terms of conjugate variables:
%......................................................
\begin{equation}
f{(\omega,k}) \sim \frac{1}{i(\omega-\omega_f)} \,\ .
\end{equation}
The response $S$ is then governed by $S=G(\omega,k)f(\omega,k)$, which has a pole at $\omega=\omega_f$ in addition to the Green's function poles.
If there are no absolute instabilities, then the $\omega$-poles are located at $\textrm{Im}(\omega_f)=0$ and possibly at 
$\textrm{Im} (\omega)<0$ if there are 
convective instabilities.

As already mentioned, in order to get the large-time behavior 
one lowers the $L_\omega$ contour. Initially the $L_\omega$ contour  is located at high $\textrm{Im} (\omega)>0$, and the
real $F_k$ contour separates upper poles $K^+$ with $\textrm{Im} (K^+)>0$  
from lower poles $K^-$  with $\textrm{Im} (K^-)<0$, where
the dispersion relation is solved in terms of  $k=K(\omega)$. The $L_\omega$ contour can be lowered down to the real axis, where
it hits the highest pole at $\omega=\omega_f$, which dominates the large-time evolution of both downstream and upstream modes
as \cite{landau}
%...............................................................
\begin{equation}
\label{eq:convsol}
S(z,t) \propto
\left \{
\begin{array}{ccc}
\exp[i K^+(\omega_f)z-i\omega_f t] & \textrm{for} & z>0\ , \\
 \exp[i K^-(\omega_f)z-i\omega_f t] & \textrm{for} & z<0\ .  \\
\end{array} \right .
\end{equation}
%...............................................................
Note that the poles $K^\pm(\omega_f)$ have been displaced from their original position. If the initial sign of $\textrm{Im}(K)$ is preserved
for both poles $K^\pm$, then the above solutions decay exponentially, both upstream and downstream, leading to a damped stable case. 
If instead at least one  $K^\pm$ pole changes sign in $\textrm{Im}(K)$  (i.e., if the pole moves into the ``wrong'' half-plane for complex $k$), as shown in the middle panel of Fig.\,\ref{fig:fincont}, then the response is exponentially amplified in at least one direction (either upstream or downstream), corresponding to a convective
instability which moves away from the source.

This argument provides a criterion to distinguish the cases
of stable damping and of convective instability: if  $\textrm{Im}(K(\omega))$ changes sign when  $\textrm{Im}(\omega)$ varies from 
$+\infty$ to $0$ at fixed  $\textrm{  Re}(\omega)=\omega_f$, then the flow is convectively unstable; if instead 
$\textrm{Im}(K(\omega))$ does not change sign, then the (stable) flow is damped.

%%%%%%%%%%%%%%%%%%%%%%%%%%%%%%%%%%%%%%%%%%%%%%%%%%%%%%%%%%%%%%%%%%%%%%%%%%%%%
\section{Two-beam case}
%%%%%%%%%%%%%%%%%%%%%%%%%%%%%%%%%%%%%%%%%%%%%%%%%%%%%%%%%%%%%%%%%%%%%%%%%%%%%%

The general formalism discussed in the previous Section is applied below to a specific case, widely considered in plasma physics
(see, e.g., \cite{{briggs,landau}}), and also suitable as a toy model for fast flavor conversions, as shown in\,\cite{Izaguirre:2016gsx}.
Namely we assume a two-beam neutrino model with a ELN spectrum
%...................................................................
\begin{equation}
G_{\bf v} = 4 \pi [ G_1 \delta({\bf v}-{\bf v}_1) + G_2 \delta({\bf v}-{\bf v}_2)] \,\ .
\label{eq:eln2mode}
\end{equation}
%...................................................................

Using this angular spectrum, one can rotate away the first term in the right-hand-side in Eq.\,(\ref{linear}) [as in  
 Eq.\,(\ref{eq:gauge})]. By assuming azimuthal symmetry with respect to the $z$-direction, and translational invariance with respect
to the transverse directions, the problem reduces to the evolution in $z$ and $t$. Under these simplifying assumptions, one gets 
a system of coupled equations for the two velocity modes
 %..............................................................................
 \begin{eqnarray}
 i \left(\frac{\partial}{\partial t} + v_1 \frac{\partial}{\partial z} \right) S_1(z,t) &=& -g_2 S_2(z,t) \,\ , \\
 i \left(\frac{\partial}{\partial t} + v_2 \frac{\partial}{\partial z} \right) S_2(z,t) &=& -g_1 S_1(z,t) \,\ ,
 \end{eqnarray}
 %..................................................................................
  where $v_1$ and $v_2$ indicate the projection of $\nu$ velocity ${\bf v}_{1,2}$ along the $z$-direction
  [see Eq.\,(\ref{eq:veloc})], and
  %.........................................................
  \begin{eqnarray}
  g_1 &=& (1-{\bf v}_1 \cdot {\bf v}_2) G_1\,\ , \nonumber \\
    g_2 &=& (1-{\bf v}_1 \cdot {\bf v}_2) G_2 \,\ .
  \end{eqnarray}
  %...........................................................
Note that these equations are equivalent to those of
of a two-level system excited by an electric field (see, e.g.,\,\cite{Haus:1979zz,Kurizki:1998dh}). If one writes explicitly these set of equations for the real and the imaginary parts of $S_1$ and $S_2$ one gets 
two sets of coupled differential equations
%..............................................................................
 \begin{eqnarray}
  \left(\frac{\partial}{\partial t} + v_1 \frac{\partial}{\partial z} \right) f_1(z,t) &=& g_2 f_2(z,t)  \,\ , \\
  \left(\frac{\partial}{\partial t} + v_2 \frac{\partial}{\partial z} \right) f_2(z,t) &=& -g_1 f_1(z,t)  \,\ ,
 \label{eq:twobeameq}
 \end{eqnarray}
 %..................................................................................
 where $(f_1,f_2)=(\textrm{Im}\,S_1,\textrm{ Re} \,S_2)$ or $(\textrm{Im} \,S_2,\textrm{ Re} \,S_1)$.

\subsection{Stability Analysis}
\label{sec:stab}

We assume the ansatz 
%................................................................
\begin{eqnarray}
f_1 &=& a_1(k,\omega) \cos(kz-\omega t) \,\ , \nonumber \\
f_2 &=& a_2(k,\omega) \sin(kz-\omega t) \,\ ,
\label{eq:ansatz}
\end{eqnarray}
%..............................................................
which implies the following dispersion relation for  $\omega$ and $k$:
%................................................
\begin{equation}
(\omega-v_1 k)(\omega-v_2 k)= \varepsilon \,\ ,
\label{eq:disptwomodes}
\end{equation}
%.....................................................
 where 
%.................................................
 \begin{equation}
  \varepsilon = g_1 g_2 \,\ , 
 \end{equation} 
%.................................................
and the amplitudes in Eq.\,(\ref{eq:ansatz}) are related by 
%..........................................
\begin{equation}
a_1(k,\omega) = \frac{g_2 a_2(k,\omega)}{\omega - v_1 k} \,\ .
\end{equation}
%............................................ 

The dispersion relation in Eq.\,(\ref{eq:disptwomodes})
can be solved either as
 %.................................................
 \begin{equation}
 \Omega(k)= \frac{1}{2}\left\{(v_1 + v_2)k \pm [k^2 (v_1-v_2)^2 + 4 \varepsilon ]^{1/2}\right\} \ ,
 \label{eq:omegabeam}
 \end{equation}
 %..................................................
 or as
%.................................................
 \begin{equation}
K(\omega)= \frac{1}{2 v_1 v_2} \left\{(v_1+v_2)\omega \pm [\omega^2 (v_1-v_2)^2 + 4\varepsilon v_1 v_2]^{1/2}\right\} \,\ .
\label{eq:kappabeam}
\end{equation}  
%............................................ 

In the absence of a coupling between the two modes ($\varepsilon=0$), Eq.\,(\ref{eq:omegabeam}) would give $\Omega_{\pm}(k)= v_{1,2} k$,
and the two modes would cross at $k=0$. Switching on the coupling ($\varepsilon\neq 0$), two different situations arise,
depending on the sign of $\varepsilon$. If $\varepsilon >0$, then $\Omega(k)$ is real for real $k$ and the system is {\em stable}.
In contrast, if $\varepsilon <0$, then $\Omega(k)$ is complex 
for $k$ in the range
%............................................................
\begin{equation}
k^2 < -4\varepsilon/(v_1-v_2)^2 \,\ ,
\label{eq:krange}
\end{equation}
%............................................................
and the system becomes {\em unstable}. 
The transition from stability to instability involves a sign change of $\varepsilon=g_1 g_2$ and thus of the relative sign
of $G_1$ and $G_2$ in Eq.\,(\ref{eq:eln2mode}).
{This  conclusion extends  to more general $G(\theta)$  where one needs a crossing from positive to negative ELN intensities
to have an instability.}

For $\varepsilon<0$, the nature of the instability (as discussed in the previous section) is determined by the imaginary part of the frequency  
$\omega_c=\Omega(k_c)$, at the wavenumber $k_c$ corresponding to null group velocity $V$. In our two-mode system $V$ reads
%-----------------
\begin{equation}
V=\frac{\partial \Omega}{\partial k} = \frac{1}{2}(v_1+v_2)\pm \frac{1}{2} \frac{k(v_1-v_2)^2}{(k^2(v_1-v_2)^2+4\varepsilon)^\frac{1}{2}}\ ,
\end{equation}
%-------------------
and vanishes for complex $k_c$ and $\omega_c$ given by  
%----------------
\begin{equation}
k_c^2 = \frac{-\varepsilon(v_1+v_2)^2}{v_1v_2(v_1-v_2)^2}\ ,
\label{eq:pinchk}
\end{equation}
%----------------- 
and 
\begin{equation}
\omega^2_c= \frac{-4\varepsilon v_1 v_2 }{(v_1-v_2)^2} \,\ ,
\label{eq:pinchom}
\end{equation}
%..........................................................
respectively.
It is easy to verify that $k_c$ is a double root of $K(\omega)$ for $\omega=\omega_c$.  
The above equations entail an unstable solution with $\textrm {Im}(\omega_c)>0$ (corresponding to  
an absolute instability) only for $v_1v_2<0$.  For unstable cases $(\varepsilon<0)$, 
the convective or absolute nature of the instability is thus determined by the sign of $v_1v_2$.
For stable cases $(\varepsilon >0)$, the function $K(\omega)$ develops an imaginary part only if $v_1v_2<0$, in which case 
the stable mode is exponentially damped. The sign of $v_1v_2$ thus distinguishes also damped and completely stable modes. 
Summarizing, one can recover the four categories introduced in Sec.\,\ref{sec:class} in terms of
the sign of $\varepsilon$ and of $v_1v_2$ as follows (see \cite{landau}).

%%%%%%%%%%%%%%%%%%%%%%%%%%%%%%%%%%%%%%%%%%%%%%%%
 \begin{figure}[!t]
\begin{centering}
\includegraphics[width=0.71\textwidth]{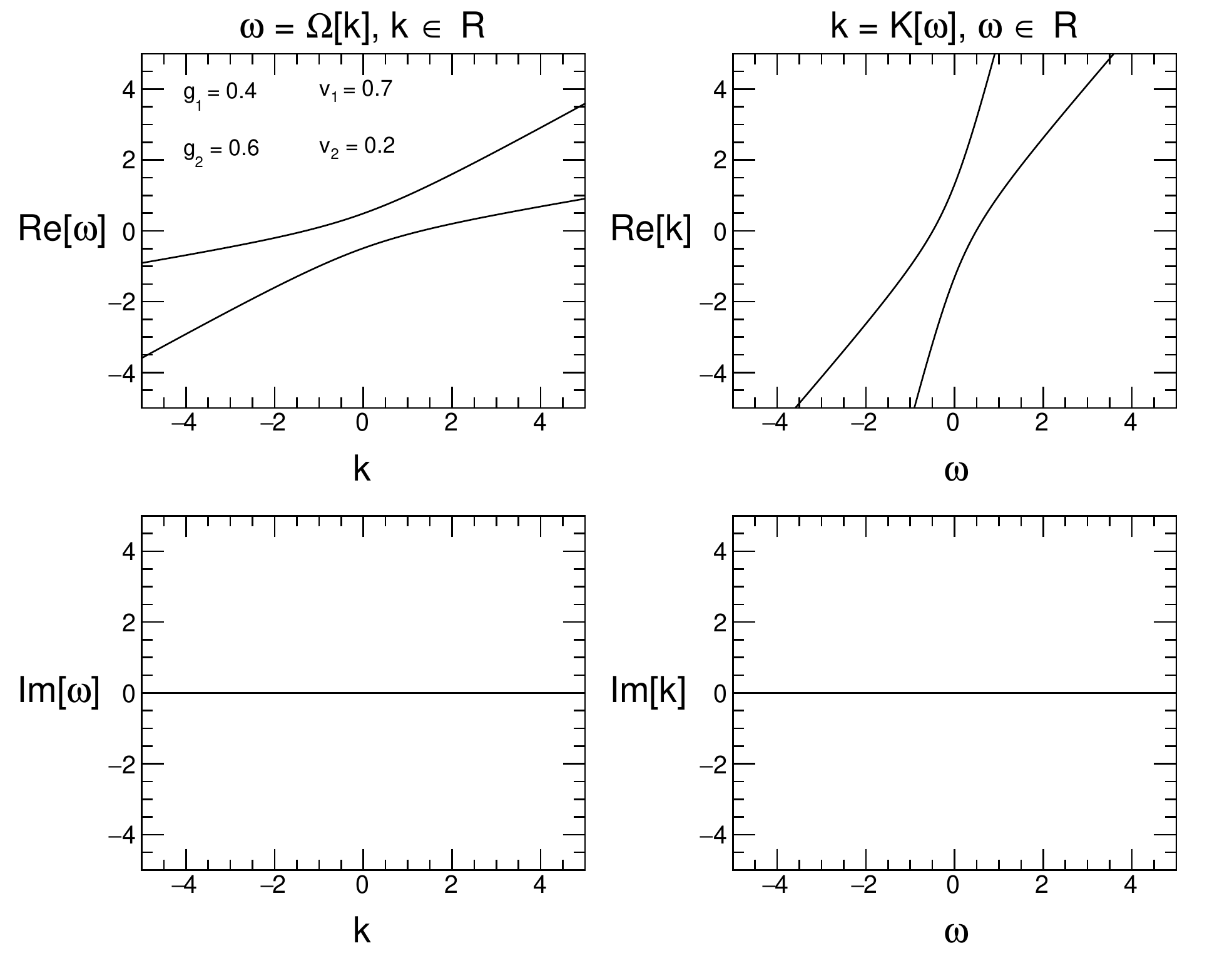}
\end{centering}
\caption{
Two-beams model: Example of solutions of the dispersion relation in a completely stable case. 
Left panels: Real part (upper panel) and imaginary
part (lower panel) of $\omega=\Omega(k)$ for real $k$ as in Eq.\,(\protect\ref{eq:omegabeam}).
Right panels:    Real part (upper panel) and imaginary
part (lower panel) of $k=K(\omega)$ for real $\omega$ as in Eq.\,(\protect\ref{eq:kappabeam}).
The two line colors indicate the two possible solutions of the dispersion relation.}
\label{fig:stab}
\end{figure} 

%%%%%
 \begin{figure}[!t]
\begin{centering}
\includegraphics[width=0.71\textwidth]{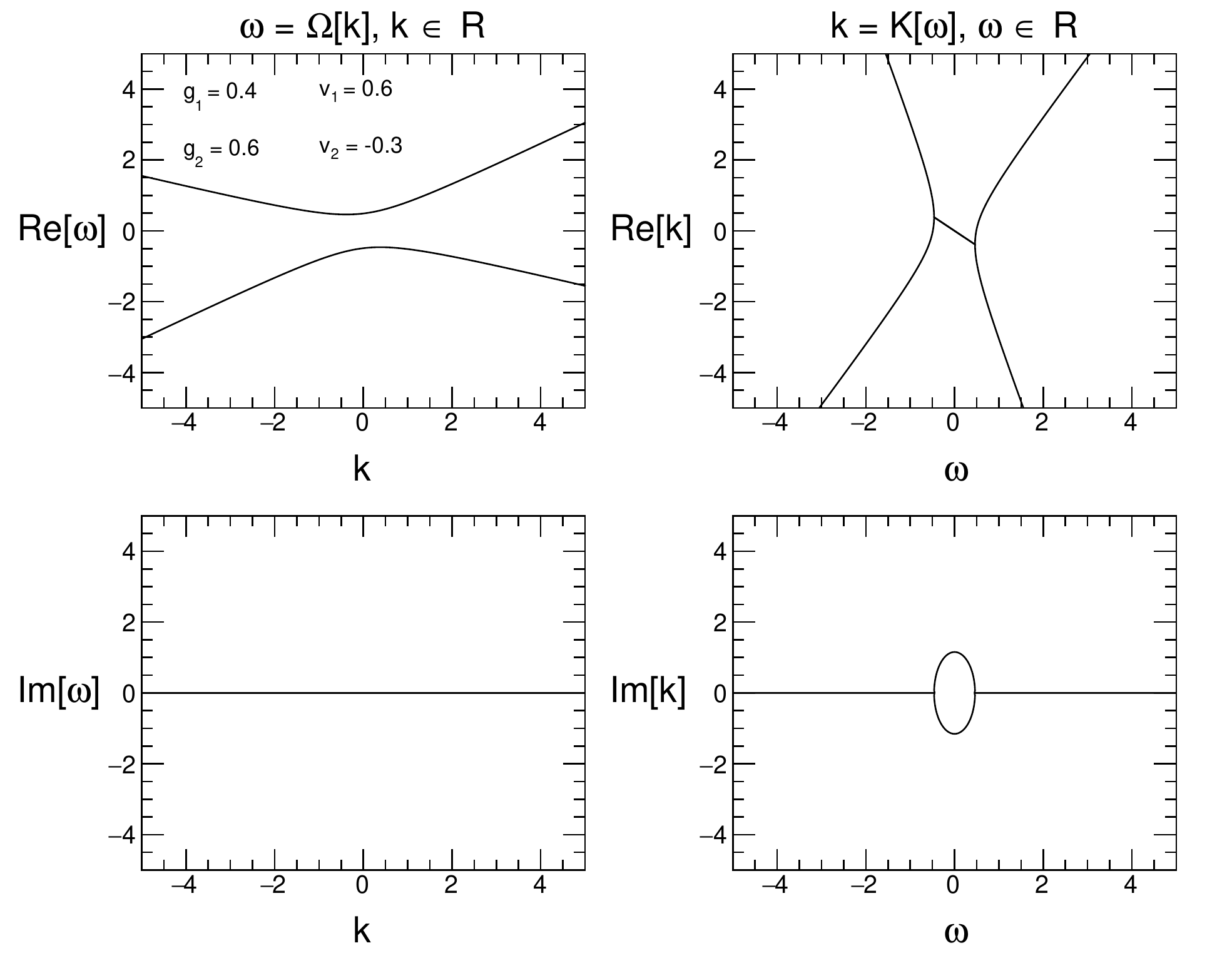}
\end{centering}
\caption{
As in Fig.\,(\protect\ref{fig:stab}), but for a damped stable case. There is gap in $\omega$, where $k$ takes complex values.}
\label{fig:nontrans}
\end{figure} 

%%%%%%%%%%%%%%%%%
\begin{figure}[!b]
\begin{centering}
\includegraphics[width=0.71\textwidth]{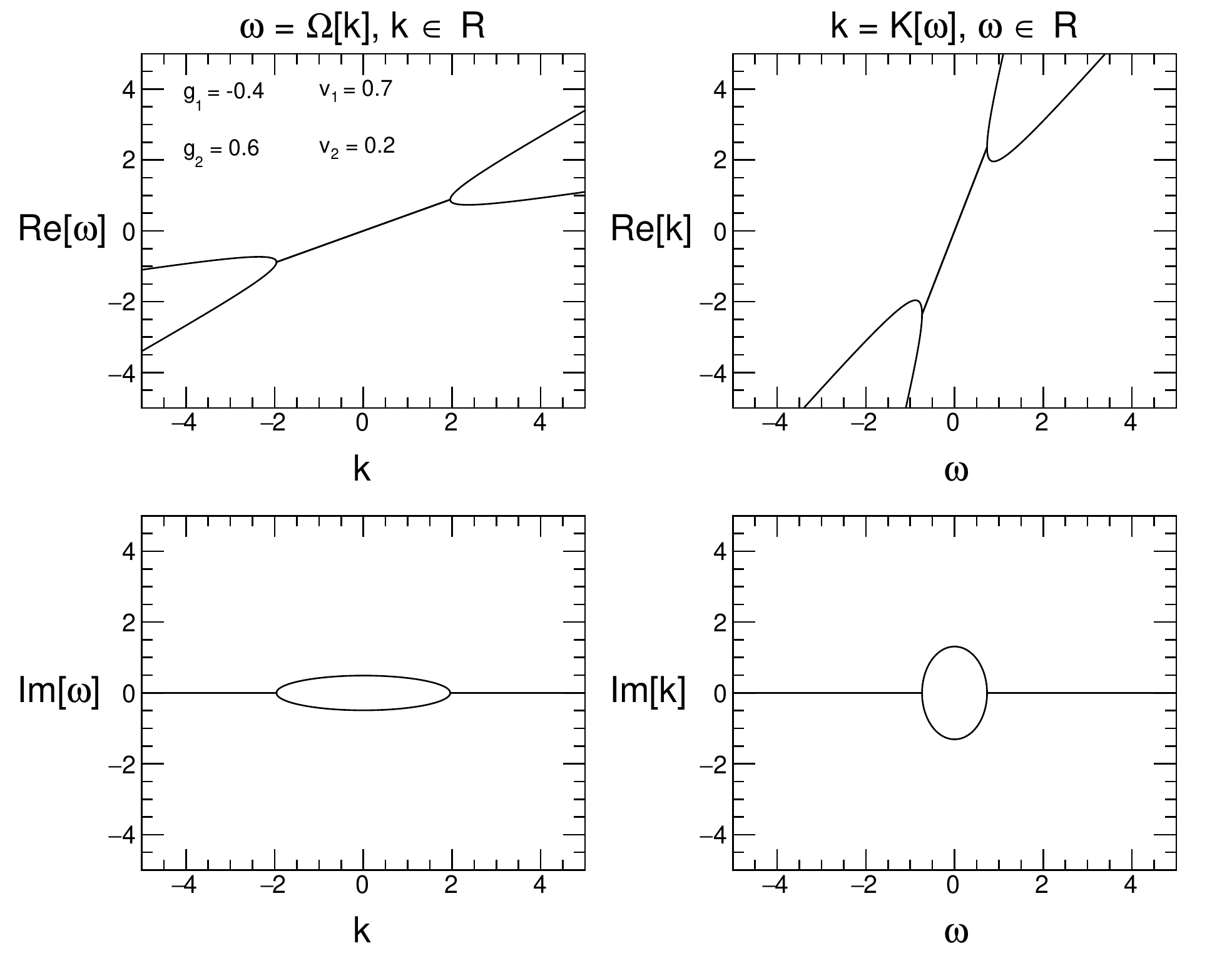}
\end{centering}
\caption{
As in Fig.\,(\protect\ref{fig:stab}), but for a convectively unstable case.
$\omega$ can take complex values for real $k$, and $k$ can take complex values for real $\omega$.}
\label{fig:convective}
\end{figure}

\begin{figure}[!t]
\begin{centering}
\includegraphics[width=0.71\textwidth]{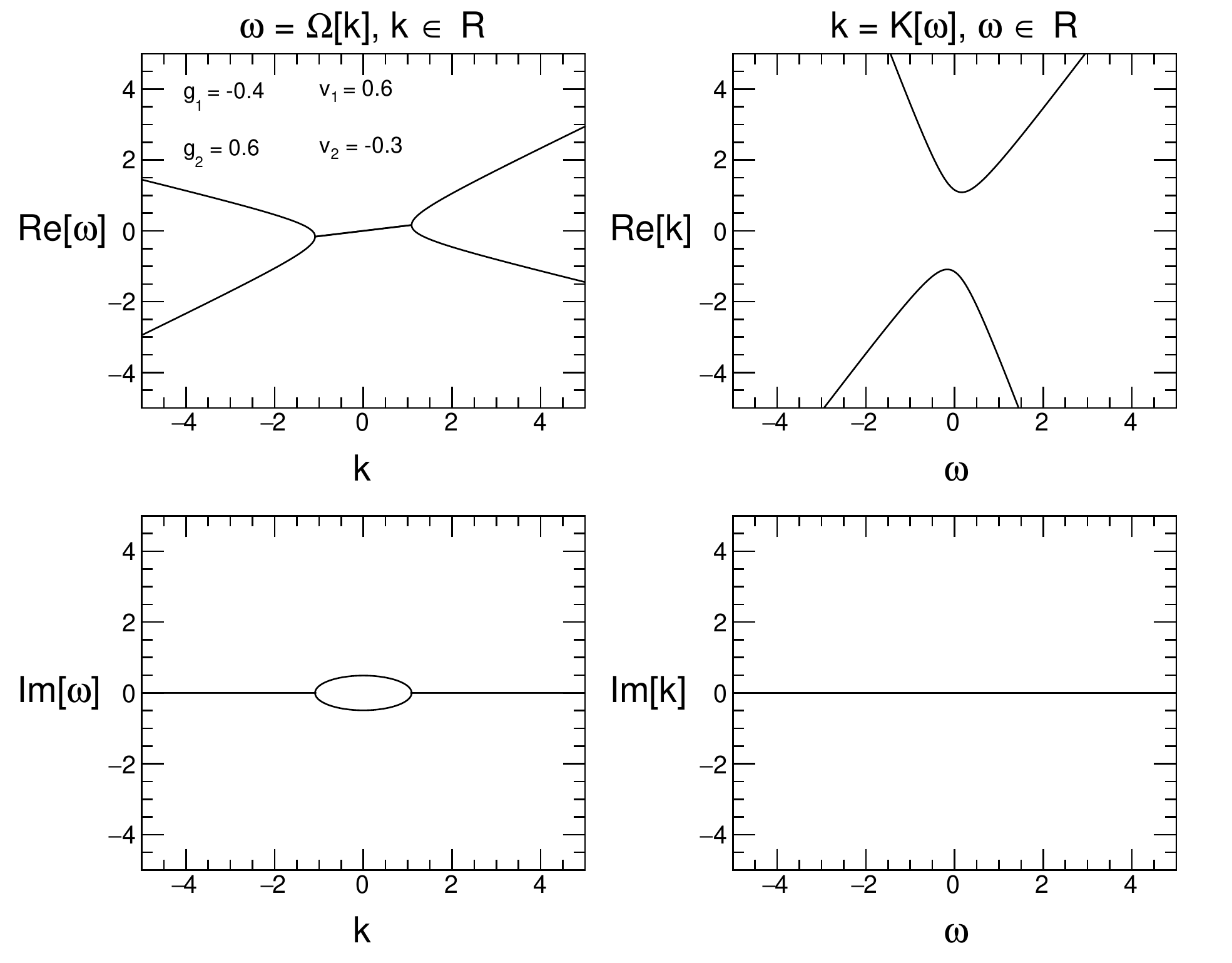}
\end{centering}
\caption{
As in Fig.\,(\protect\ref{fig:stab}), but for an absolutely unstable case. There is gap in $k$, where $\omega$ takes complex values.}
\label{fig:abs}
\end{figure}

\begin{itemize}

\item[$\bullet$] 
\emph{Complete stability} ($\varepsilon >0$, $v_1 v_2>0$), see Fig.\,\ref{fig:stab}. 
In this case $\Omega(k)$ is real for real $k$ and $K(\omega)$ is real  for real $\omega$. No imaginary part is developed,
 and the system is completely stable.

\item[$\bullet$]

\emph{Stability with damping} ($\varepsilon >0$, $v_1 v_2<0$), see Fig.\,\ref{fig:nontrans}. The function $\Omega(k)$ is real for real $k$, and the system
cannot exhibit convective or absolute instability. However,  $K(\omega)$ is complex in the frequency range
%............................................................
\begin{equation}
\omega^2 <   \frac{-4\varepsilon v_1 v_2 }{(v_1-v_2)^2} \,\ , 
\label{eq:rangeom}
\end{equation}
%..........................................................
graphically corresponding to a ``gap'' in $\omega$ ,
where perturbations decay exponentially in the space coordinate $z$.

\item[$\bullet$]
\emph{Convective instability} ($\varepsilon <0$, $v_1 v_2>0$), see Fig.\,\ref{fig:convective}.
In this case the dispersion relation can provide
{complex $\omega$ for real $k$ and complex $k$ for real $\omega$}.
Following the criteria outlined in Sec.\,\ref{sec:briggs}, we consider the path of the complex roots  
$K(\omega)$ of Eq.\,(\ref{eq:kappabeam}) from large to small $\rm{Im}(\omega)$, assuming $v_{1,2}>0$ for definiteness. 
For $\rm{Im}(\omega)\to\infty$, both roots are in the same upper half of the complex plane,
\begin{equation}
k \approx \omega/v_1 \,\ ,  \,\ \,\ \,\ \,\ \,\ k \approx \omega/v_2 \,\ .
\label{eq:kasi}
\end{equation}
When $\textrm{Im}(\omega)\to 0$, the roots $K(\omega)$ in the range of 
of Eq.\,(\ref{eq:rangeom}) are instead complex conjugate, and the one for which 
$\textrm{Im} (K(\omega))<0$ has migrated from the upper to the lower half of the complex plane. 
Therefore, in the range of Eq.\,(\ref{eq:rangeom}) we have a convective instability
propagating in the positive $z$ direction [see Eq.\,(\ref{eq:convsol})].

%%%%%%%%%%%%%%%%%%%%%%%%%%%%%%%%%%%%%%%%
\item[$\bullet$]
\emph{Absolute instability} ($\varepsilon <0$, $v_1 v_2<0$), see Fig.\,\ref{fig:abs}.
In this case 
the function $K(\omega)$ is real for all real $\omega$, but the function $\Omega(k)$ is complex
in the range of Eq.\,(\ref{eq:krange}), corresponding to a gap in $k$.
We note from Eq.\,(\ref{eq:kasi}) that for $\textrm{Im}(\omega) \to \infty$, since $v_1 v_2 <0$
the two roots $K(\omega)$ are in opposite half-planes. Such roots  coalesce 
at a single pinching point as $\textrm{Im}(k)$ is gradually lowered, down to 
imaginary $\omega_c$ value derived from Eq.\,(\ref{eq:pinchom}).

\end{itemize}

%%%%%%%%%%%%%%%%%%%%%%%%%%%%%%

\begin{figure}[!t]
\begin{centering}
\includegraphics[width=0.8\textwidth]{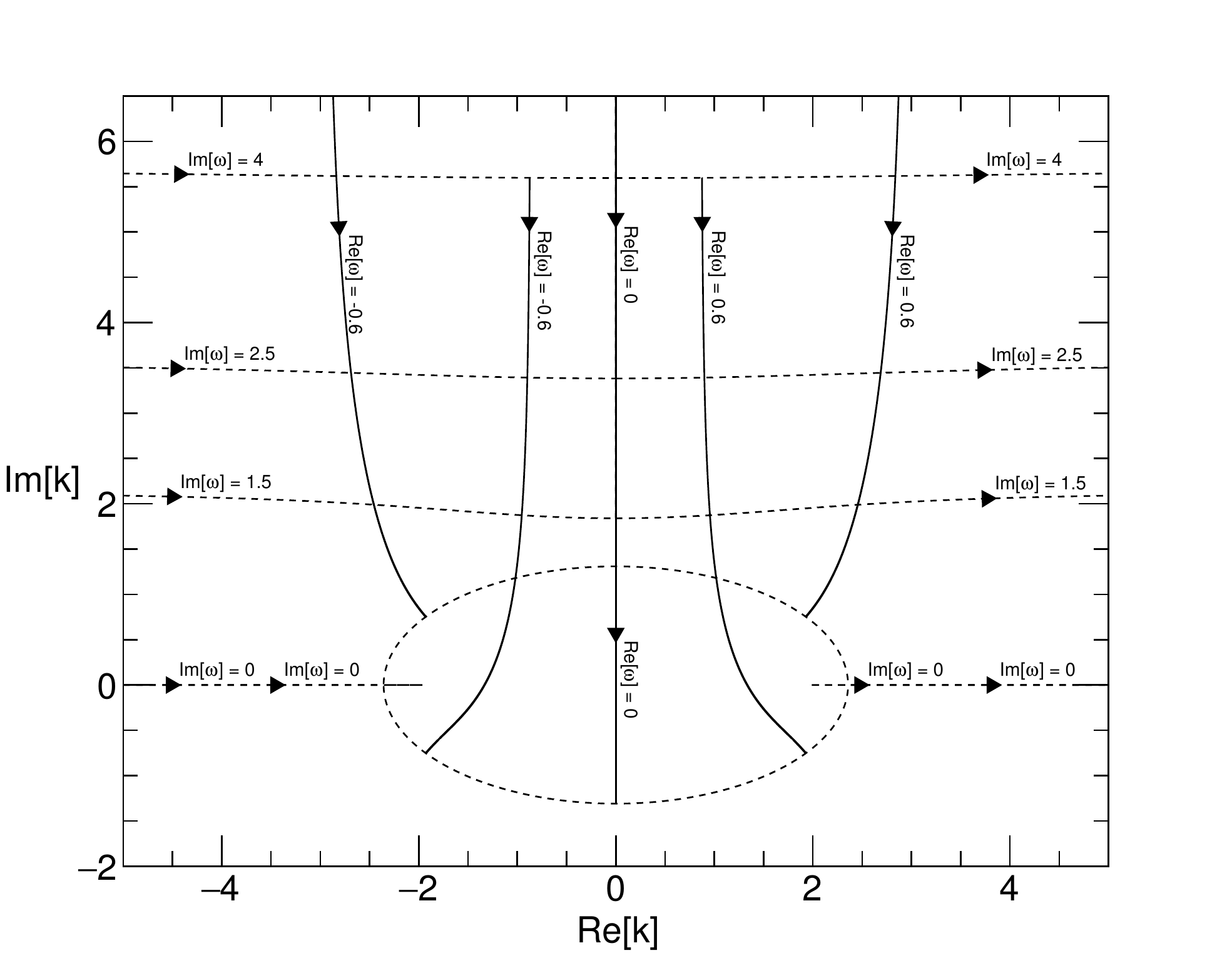}
\end{centering}
%\vspace{0.5cm}
\caption{
Two-beams model:  Convective instability in the $k$-plane, for the same 
two-mode parameters as in Fig\,\protect\ref{fig:convective}. Nearly vertical and solid lines indicate $K(\omega)$ roots of the dispersion
relation for the indicated value of $\textrm{Re}(\omega)$.  Nearly horizontal and dashed lines indicate isocontours of $\textrm{Im}(\omega)$.
The two colors refer to the two roots of the dispersion relation. Lowering $\textrm{Im}(\omega)$ to zero at fixed $\textrm{Re}(\omega)$, one 
of the two complex roots migrates from the upper to the lower half-plane, signaling a convective instability.}
\label{fig:conv}
\end{figure}

\begin{figure}[!t]
\begin{centering}
\includegraphics[width=0.8\textwidth]{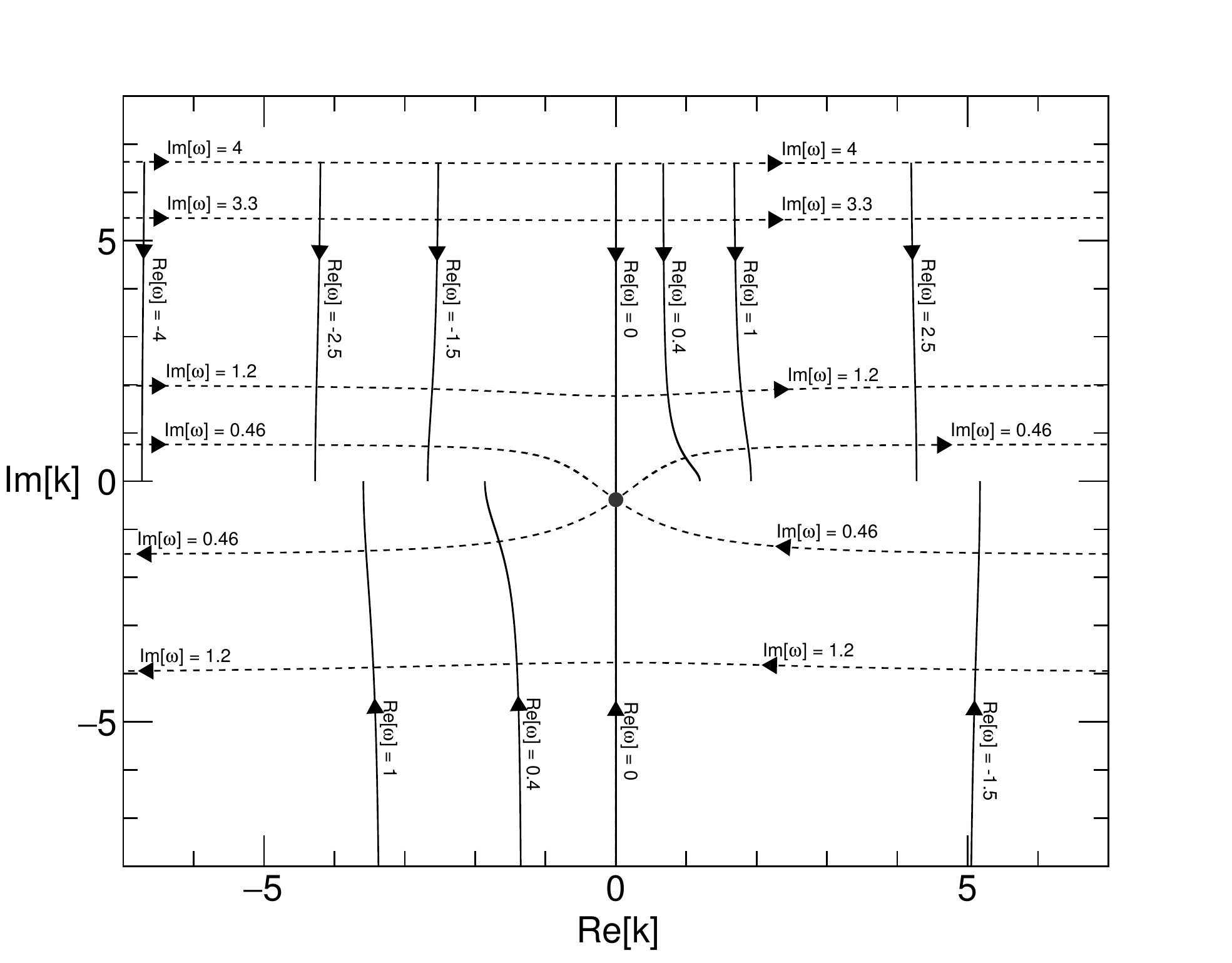}
\end{centering}
%\vspace{-0.5cm}
\caption{
Two-beams model:  Absolute instability in the $k$-plane, for the
same two-mode parameters as in Fig.\,\protect\ref{fig:abs}. Nearly vertical and solid lines indicate the two roots $K^\pm(\omega)$ (distinguished by color) 
of the dispersion relation for fixed $\textrm{Re}(\omega)$. Nearly horizontal and dashed lines indicate isocontours of $\textrm{Im}(\omega)$. 
Lowering $\textrm{Im}(\omega)$ eventually leads to a pinching of the two roots, denoted by the dot.}
\label{fig:pinch}
\end{figure}

The behavior of the $K(\omega)$ roots in the complex $k$ plane for convective and absolute instabilities is further illustrated 
in Figs.\,\ref{fig:conv} and\,\ref{fig:pinch}, respectively.
In  Fig.\,\ref{fig:conv},  nearly
 vertical and solid curves correspond to $K(\omega)$  roots of the dispersion
relation for the indicated value of $\textrm{Re}(\omega)$, while nearly horizontal and dashed curves correspond to isocontours of $\textrm{Im}(\omega)$.
The two colors refer to the two roots of the dispersion relation.
Lowering $\textrm{Im}(\omega)$ towards zero, one of the two solutions at given $\textrm{Re}(\omega)$ migrates from upper to lower plane
and gives rise to a convective instability.
In Fig.\,\ref{fig:pinch}, nearly vertical lines indicate $K_+(\omega)$ and $K_-(\omega)$ roots of the dispersion
relation for the indicated value of $\textrm{Re}(\omega)$. Horizontal dashed lines indicate isocontours of $\textrm{Im}(\omega)$. 
By lowering $\mathrm{Im}(\omega)$, the pinching point is eventually found
at $\mathrm{Im}(\omega_c) = 0.46$ and $\mathrm{Im}(k_c)=-0.38$,  in agreement with Eqs.\,(\ref{eq:pinchk})  and (\ref{eq:pinchom})
for the chosen two-mode parameters.

%...............................................................

%%%%%%%%%%%%%%%%%%%

We conclude this section with some comments on the growth rate along a ray $V=z/t$ for unstable flows ($\varepsilon<0$), as given in Eq.\,(\ref{eq:sigma}). For
the two-mode system we find:
\begin{equation}
\label{eq:growth}
\sigma = |\varepsilon|^\frac{1}{2} \left[\frac{(v_1-v_2)^2-\delta^2}{(v_1-v_2)^2}\right]^\frac{1}{2}\ ,
\end{equation} 
where 
\begin{equation}
\delta= 2V - (v_1+v_2)\ .
\label{eq:delta}
\end{equation}
The maximum growth rate $\sigma=\sqrt{\varepsilon}$ occurs at $\delta=0$, namely, $V=(v_1+v_2)/2$, which applies to both convective and absolute 
instabilities. Only for $v_1v_2<0$ (absolutely unstable case) there growth rate is also defined at $V=0$, and one finds then
$\sigma = \textrm{Im}(\omega_c)$ with $\omega^2_c$ as in Eq.\,(\ref{eq:pinchom}). Finally, notice that the growth rate vanishes at $V=v_{1,2}$.

%%%%%%%%%%%%%%%%%%%%%%%%%%%%%%%%%%%%%%%%%%%%%%%%%%%%%%%%%%%%%%%%%%%%%%%%%%%%%%%%%%%%%%%
\begin{figure}
\centering
\includegraphics[width=0.45\columnwidth]{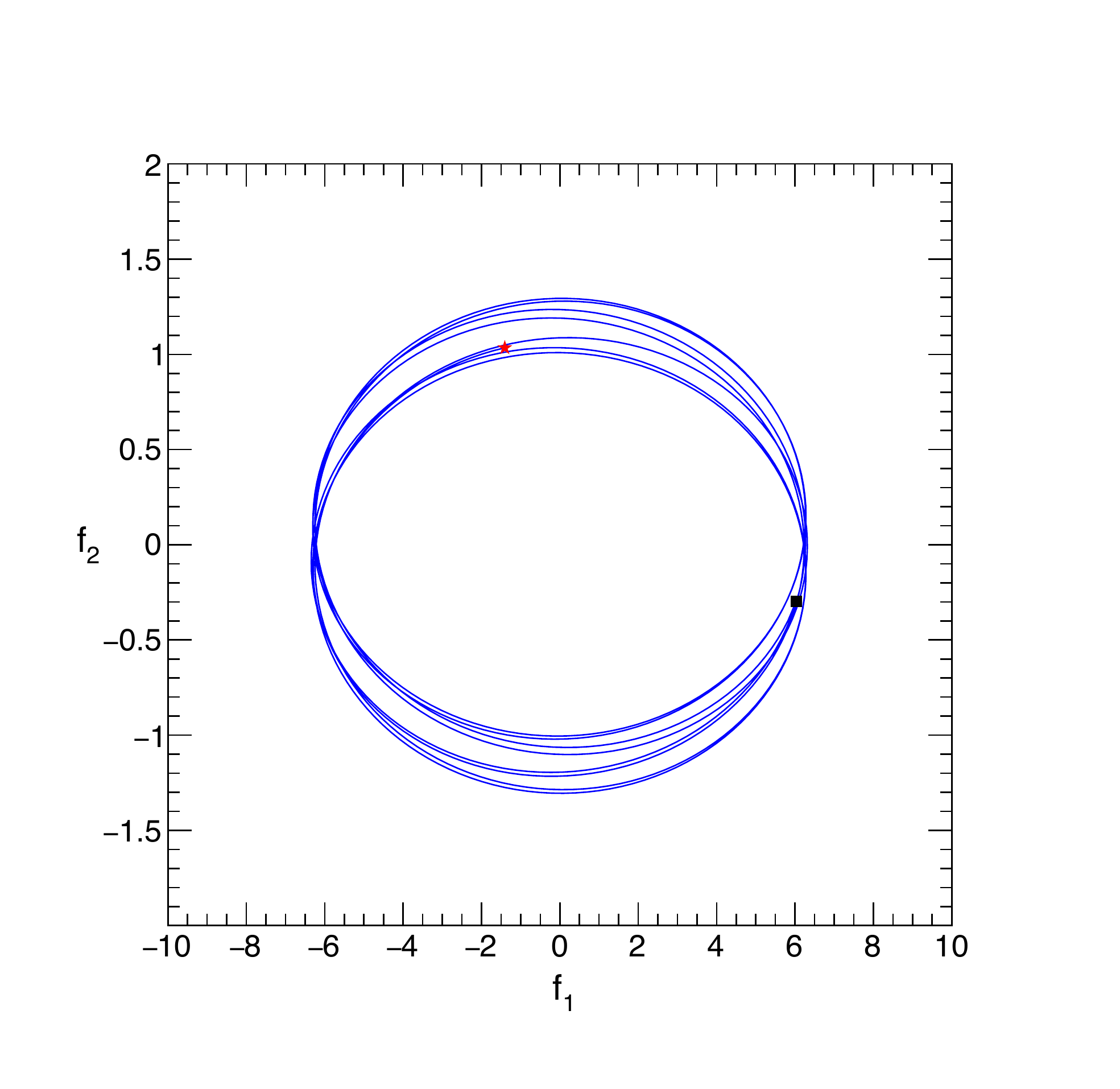}
\hspace{-1.cm}
\includegraphics[width=0.45\columnwidth]{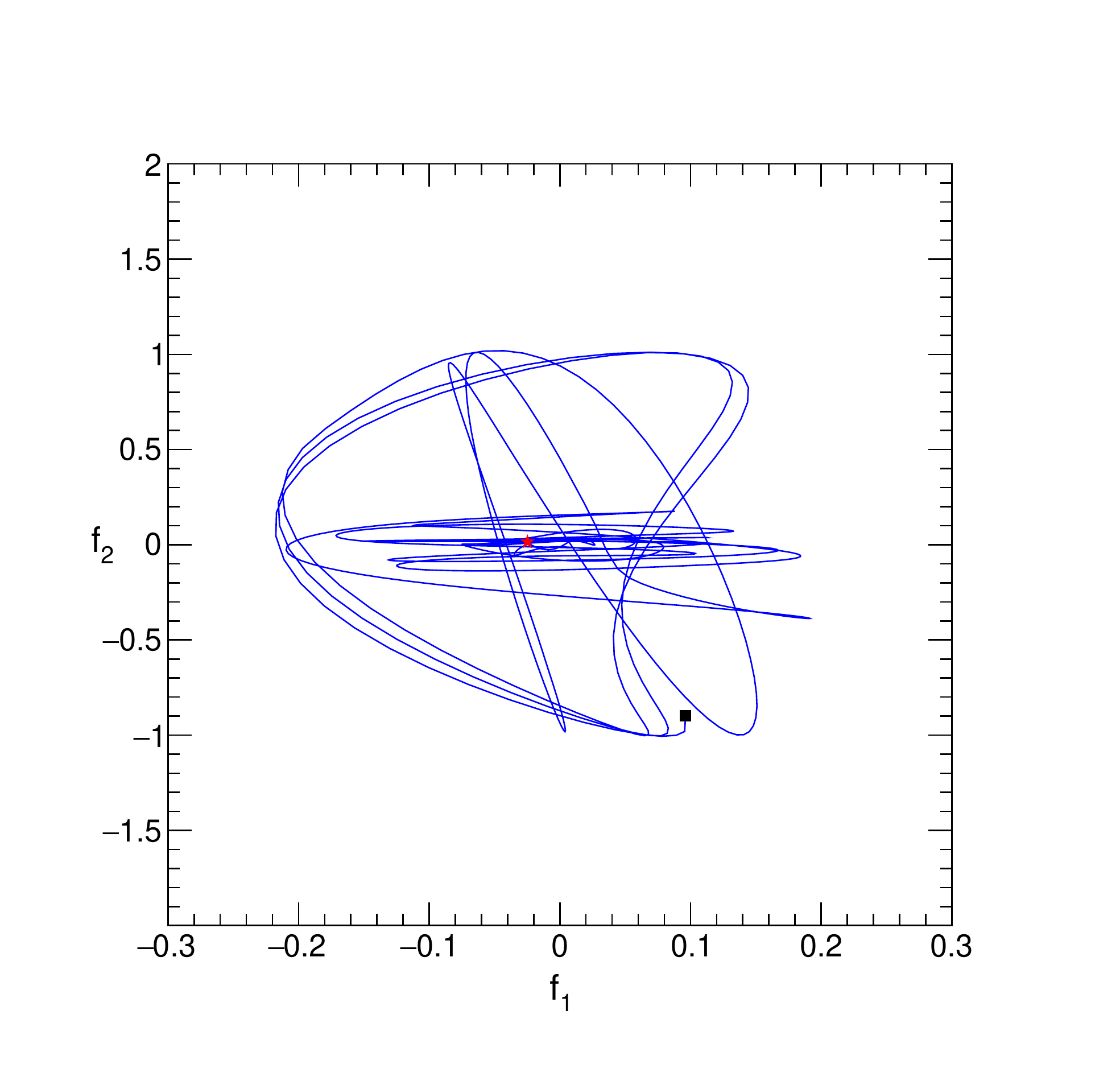}\\
\vspace{-1.cm}
\includegraphics[width=0.45\columnwidth]{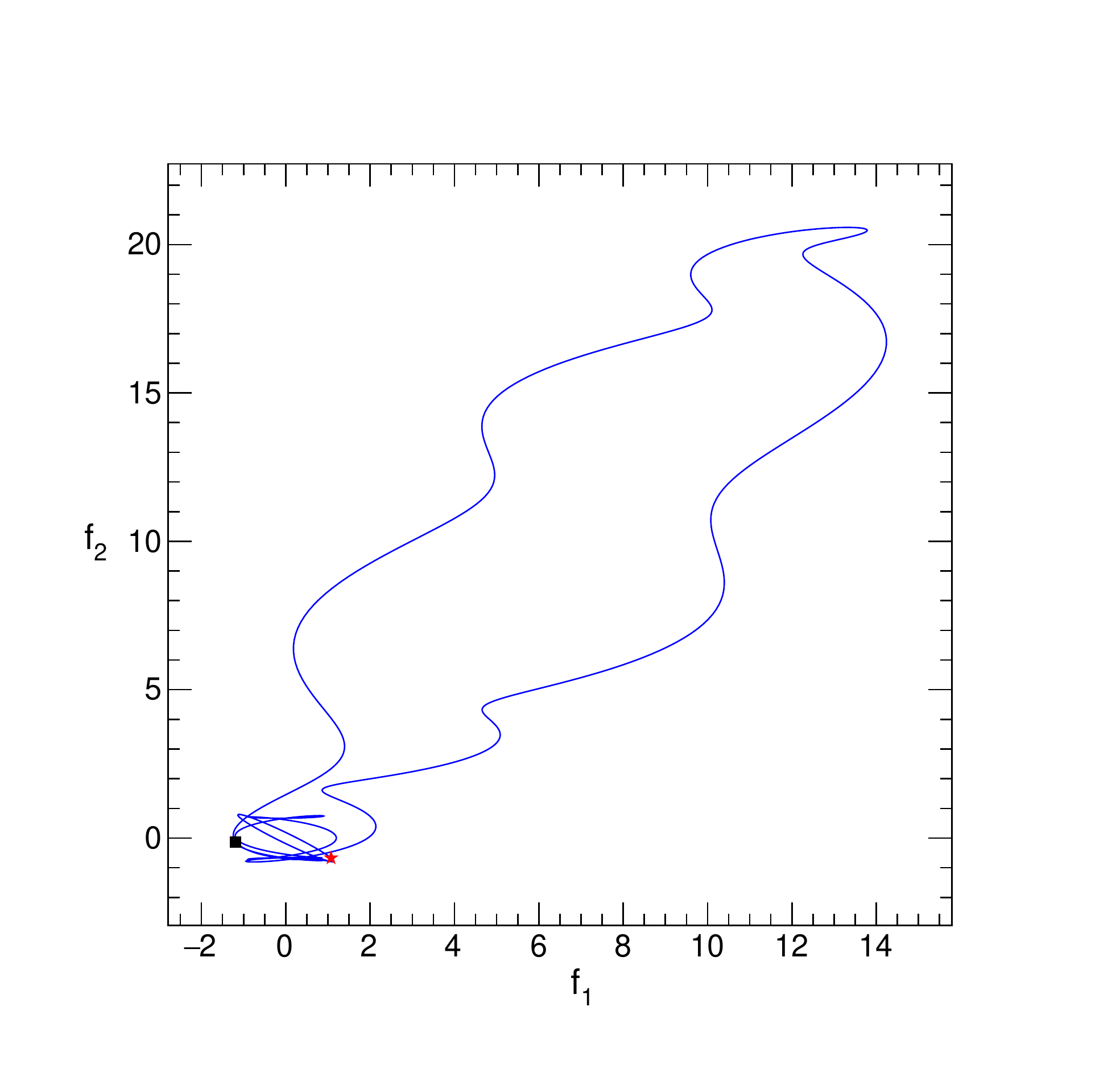}
\hspace{-1.cm}
\includegraphics[width=0.45\columnwidth]{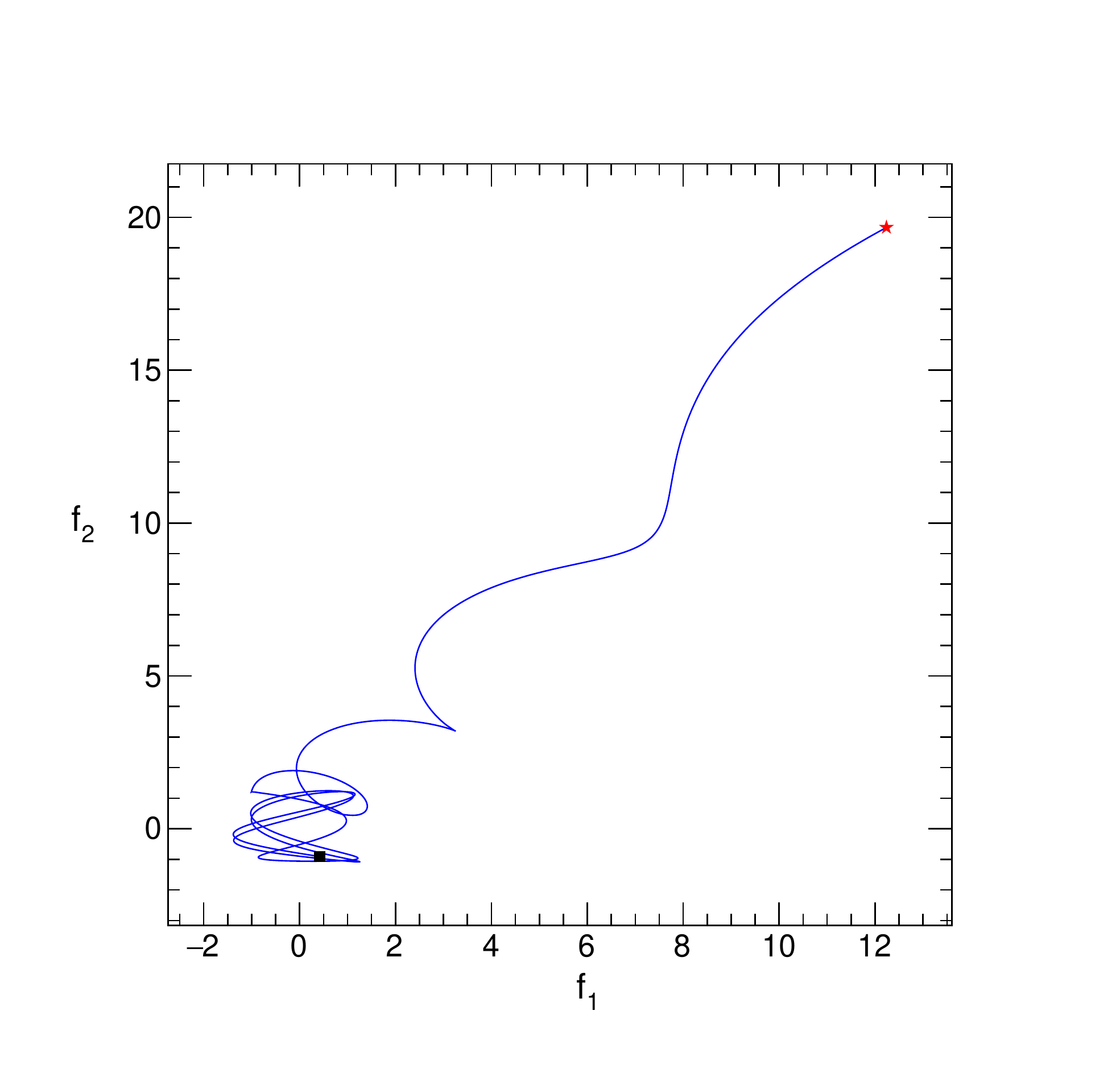}
\vspace{-.5cm}
\caption{
Orbits of plane-wave solutions of Eq.\,(\ref{eq:ansatz}) in the plane $(f_1,\, f_2)$ 
at fixed $z$  function of time $t$, in the presence of small perturbations.
Left upper panel: completely stable case ($g_1=0.4$, $g_2=0.6$, $v_1=0.7$, $v_2=0.2$). Right upper panel: damped stable case
($g_1=0.4$, $g_2=0.6$, $v_1=0.6$, $v_2=-0.3$).  Left lower  panel: convectively unstable case ($g_1=-0.4$, $g_2=0.6$, $v_1=0.7$, $v_2=0.2$).
Right lower  panel: absolutely  unstable case ($g_1=-0.4$, $g_2=0.6$, $v_1=0.6$, $v_2=-0.3$).
The square marker indicates the solution at $t=0$ and the star marker the position at large final $t$. Note the change in scale for lower panels.} 
\label{fig:orbit}
\end{figure}
%%%%%%%%%%%%%%%%%%%%%%%%%%%%%%%%%%%%%%%%%%%%%%%%%%%%%%%%%%%%%%%%%%%%%%%%%%%

\subsection{Particle-like vs Tachyon-like Dispersion}

Here, we briefly mention another possible interpretation of the instabilities discussed above.
We consider the set of the coupled differential equations  [Eq.\,(\ref{eq:twobeameq})] for the two-beam problem
{transforming the time coordinate as\,\cite{poladian,longhi}
%......................................
\begin{equation}
t^{\prime}= t - \frac{1}{2}\left[\frac{1}{v_1}+ \frac{1}{v_2} \right] z \,\ ,
\end{equation}
%..............................
so that the apparent group velocity with the new time coordinate are the same in magnitude  but opposite in sign
and given by
%............................
\begin{equation}
\frac{1}{{\bar c}} = \frac{1}{2} \left[\frac{1}{v_1}-\frac{1}{v_2} \right] \,\ .
\end{equation}
%.....................................
In this reference  system one obtains as coupled equations
%..............................................................................
 \begin{eqnarray}
  \left(\frac{\partial}{\partial z} + \frac{1}{{\bar c}}\frac{\partial}{\partial t^\prime} \right) f_1(z,t^\prime) &=& \frac{g_2}{v_1} f_2(z,t^\prime)  \,\ , \\
 \left(\frac{\partial}{\partial z} - \frac{1}{{\bar c}}\frac{\partial}{\partial t^\prime}  \right) f_2(z,t^\prime) &=& -\frac{g_1}{v_2} f_1(z,t^\prime)  \,\ ,
 \label{eq:twobeameq1}
 \end{eqnarray}
 %..................................................................................
Applying to
the first equation of the system in Eq.\,(\ref{eq:twobeameq1})
the differential operator of the second equation (or viceversa)
one obtains the  Klein-Gordon equation 
%%%%%%%%%%%%%%%%%%%%%%%%%%%%%%%%%%%%%%%%%%%%%
\footnote{Note that in the moving frame, one can introduce a spinor wave field $\psi= (f_1,f_2)^T$, writing
Eq.\,(\ref{eq:twobeameq1}) as a one-dimensional Dirac equation\,\cite{longhi}.}
%
%%%%%%%%%%%%%%%%%%%%%%%%
  %....................................................
\begin{equation}
 \left(\frac{1}{{\bar c}^2}\frac{\partial^2}{\partial t^2} -  \frac{\partial^2}{\partial z^2} + m^2 {\bar c}^2 \right) f_{1,2}= 0 \,\ ,
 \label{eq:waveeq}
\end{equation}
 %...................................................
 with 
 %.............................................
\begin{equation}
	m^2 {\bar c}^2= -\frac{\varepsilon}{v_1 v_2} \,\ .
\end{equation}
 %.............................................
From  Eq.\,(\ref{eq:waveeq})
 one obtains as dispersion relation
 %.....................................
 \begin{equation}
 \frac{\omega^2}{{\bar c}^2} -k^2= m^2 {\bar c}^2 \,\ .
 \label{eq:dispequad}
 \end{equation}
 %...............................
If $m^2>0$, this quantity plays the role of a mass term, 
 then Eq.\,(\ref{eq:dispequad}) is the dispersion relation of a particle, having a gap in $\omega$.
From the dispersion relation, one realizes that if $\omega^2> m^2 {\bar c}^4$, Eq.\,(\ref{eq:waveeq}) would have oscillatory waves as the solution, while 
if $\omega <m^2 {\bar c}^4$ it would represent damped oscillatory waves.
This is consistent with what found for the in the damped  case in the previous Section for two counter-propagating modes with
$\varepsilon >0$.

If we now move to the case of $ m^2 <0$, Eq.\,(\ref{eq:waveeq}) would represent a Klein-Gordon equation
with imaginary mass. 
In this case the dispersion relation of Eq.\,(\ref{eq:dispequad}), with a gap in $k$, would be the one expected for ``tachyons''\,\cite{Aharonov:1969vu,Andreev:1996qs}.
From this dispersion relation if $k^2> m^2 {\bar c}^2$ one would expect normal oscillatory motion. Conversely,
for  $k^2< m^2 {\bar c}^2$ we would have an exponential growing solution. It is intriguing to realize that the absolute instability found for two counter-propagating modes with $\varepsilon <0$
 is of  a ``tachyonic'' type.

%%%%%%%%%%%%%%%%%%%%%
\begin{figure}[!t]
\begin{centering}
\includegraphics[width=0.8\textwidth]{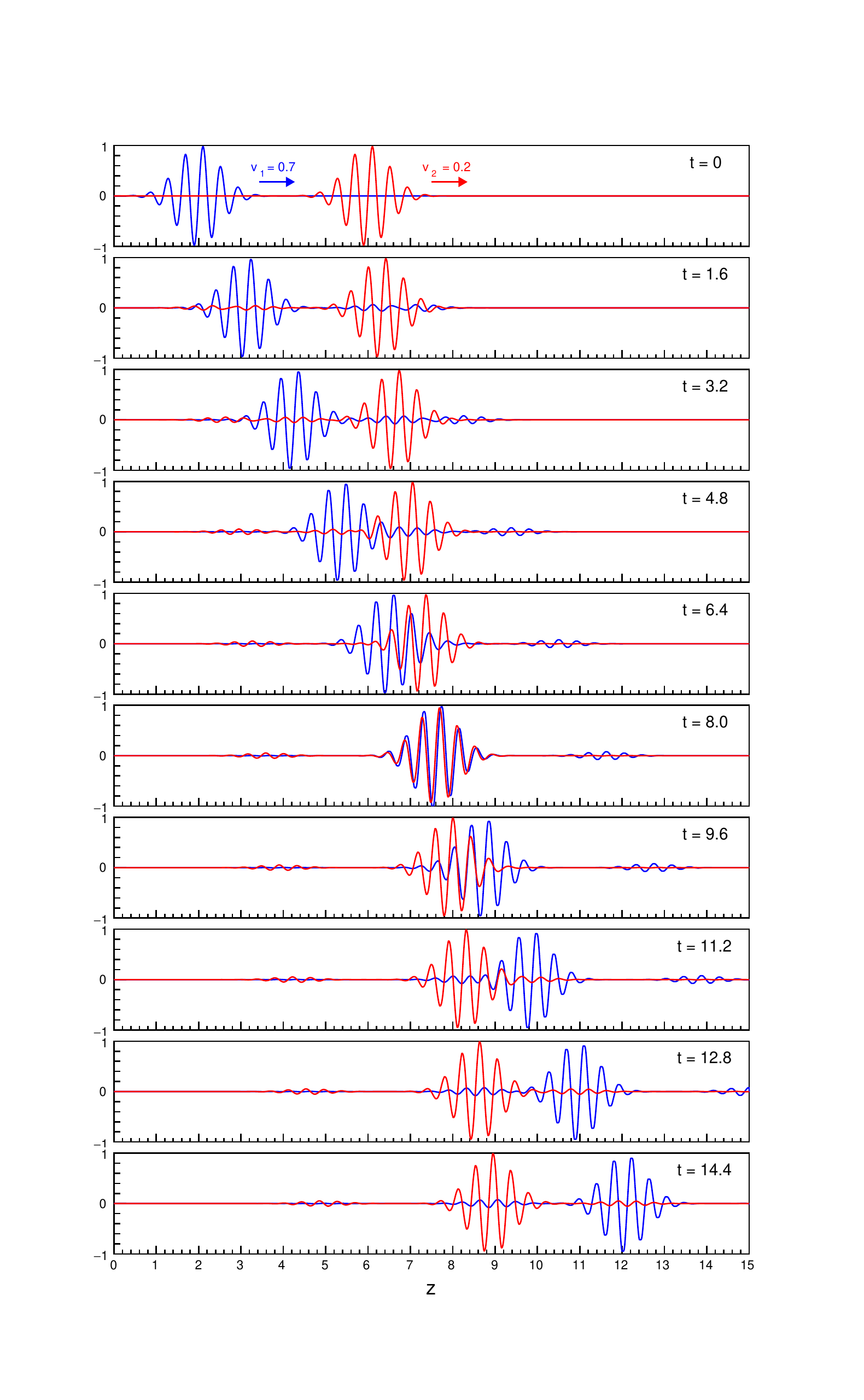}
\end{centering}
\vspace{-1.cm}
\caption{Two-beams model with traveling wavepackets: Completely stable case with $g_1=0.4$, $g_2=0.6$, $v_1=0.7$, $v_2=0.2$.}
\label{fig:stab_num}
\end{figure} 
%%%%%%%%%%%%%

\begin{figure}[!t]
\begin{centering}
\includegraphics[width=0.8\textwidth]{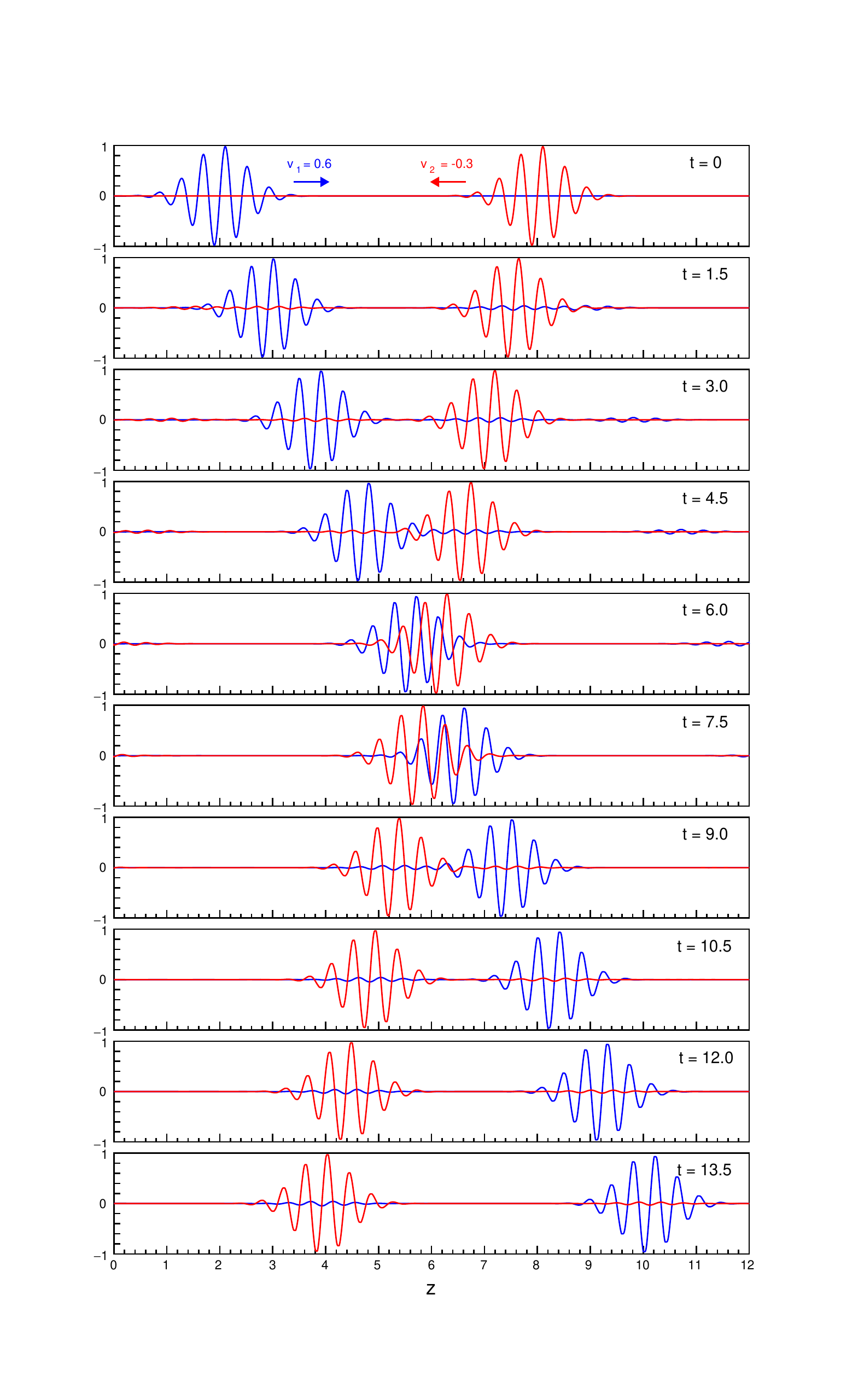}
\end{centering}
\vspace{-1.cm}
\caption{Two-beams model with traveling wavepackets: Damped case with $g_1=0.4$, $g_2=0.6$, $v_1=0.6$, $v_2=-0.3$. {The damping of perturbations is small and not apparent here.}}
\label{fig:nontransp_num}
\end{figure} 

%%%%%%%%%%%%%%%%%%%%%%%

\begin{figure}[!t]
\begin{centering}
\includegraphics[width=0.8\textwidth]{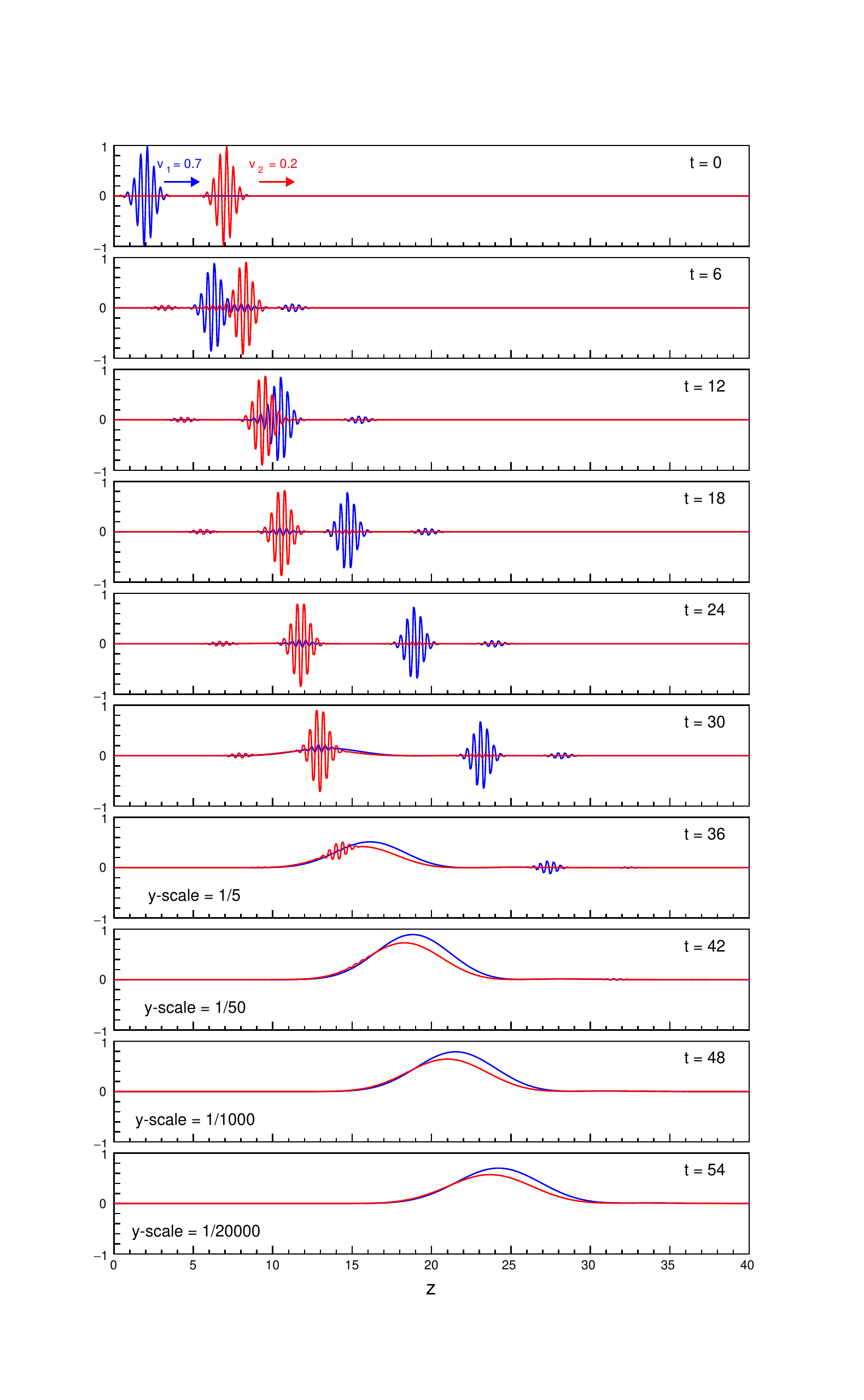}
\end{centering}
\vspace{-1.cm}
\caption{Two-beams model with traveling wavepackets: Growth of a convective instability with $g_1=-0.4$, $g_2=0.6$, $v_1=0.7$, $v_2=0.2$.
In the lower five panels, the $y$-axis is rescaled by the indicated factors, in order to display the exponentially growing disturbances.}
\label{fig:conv_num}
\end{figure} 

%%%%%%%%%%%%%

\begin{figure}[!t]
\begin{centering}
\includegraphics[width=0.8\textwidth]{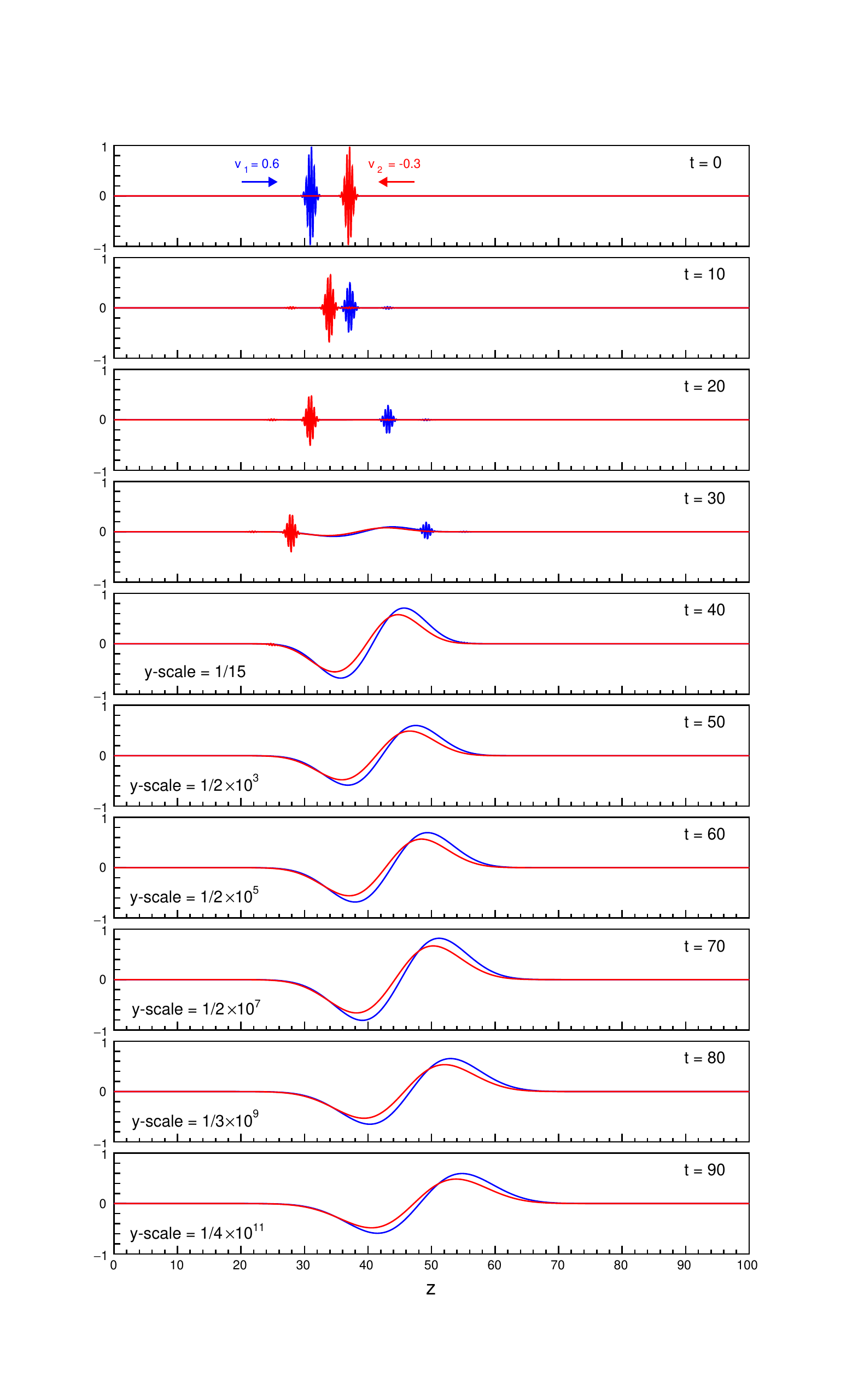}
\end{centering}
\vspace{-1.cm}
\caption{Two-beams model with traveling wavepackets: Growth of an absolute instability with $g_1=-0.4$, $g_2=0.6$, $v_1=0.6$, $v_2=-0.3$.
In the lower six panels, the $y$-axis is rescaled by the indicated factors, in order to display the exponentially growing disturbances.}
\label{fig:abs_num}
\end{figure} 

%\newpage

\subsection{Numerical Results}

In order to illustrate the predictions of the stability analysis for the four cases 
discussed in the previous Sections, we work out representative  numerical solutions of 
Eq.\,(\ref{eq:twobeameq}). 
We assume a length interval $z\in [0,\,L]$.
{In the following we will work in the units in which the neutrino potential in Eq.\,(\ref{eq:dens}) is $\mu=1$. Therefore
times and length are expressed in units of $\mu^{-1}$. Moreover, we will assume ${\mathcal O}(1)$ initial values for $f_1$ and $f_2$, since in the linear regime
these are just  arbitrary normalization factors.}

 When $v_1v_2>0$ we assume that the two
modes are emitted   at $z=0$. Conversely when  $v_1v_2<0$ we assume that the mode
with $v_1>0$ is emitted at $z=0$, while the mode with $v_2<0$ is emitted at $z=L$. 
 
At first, we look for plane-wave solutions, given by Eq.\,(\ref{eq:ansatz}). 
We assume the same numerical parameters as for the four cases in Sec.\,\ref{sec:stab}.
In Fig.\,\ref{fig:orbit} we plot the orbits of the solution ($f_1,\,f_2$) as a function of time $t$ 
at some fixed $z$ (the exact value being irrelevant for our discussion). To orbits start at the square point (at $t=0$) 
and terminate at the star point (at some large $t$). If  $\omega$ and $k$ were taken both real and exactly related by the dispersion relation, 
the orbits would be stable ellipses (not shown). We instead allow small numerical deviations from the
dispersion relation, in order to mimic perturbations and to study their evolution. 
In the completely stable case ($v_1v_2>0$ and $\varepsilon >0$, upper left panel), perturbations are neither enhanced nor suppressed, and the orbits
remain quite stable in time.   
In the stable case with damping ($v_1v_2<0$ and $\varepsilon >0$, upper right panel), perturbations
eventually lead to a decay of the solution amplitudes, signaled by the shrinking of the orbit towards $f_1=f_2=0$. 
In the convectively unstable case ($v_1v_2>0$ and $\varepsilon <0$, lower left panel), perturbations are amplified and orbits diverge for a transient time, until the unstable flow
moves away and decays at any fixed $z$. 
 Finally, in the case of absolute instability  ($v_1v_2<0$ and $\varepsilon <0$, lower right panel), after a transient the orbit parameters tend to 
grow indefinitely with time.

In order to go beyond monochromatic waves, we perform a second numerical experiment by launching
two  wavepackets with different velocities $v_1$ and $v_2$, with an initial shape
%........................................................
\begin{equation}
f_{1,2}(z,t=0) = \frac{1}{\sqrt{2 \pi \sigma_f^2}}\exp\left[ -\frac{(z-z_{1,2})^2}{2 \sigma_f^2} \right] \sin k (z-z_{1,2}) \,\ ,
\end{equation}
%...................................................
Numerically we fix $k=15$, $\sigma_f=0.5$, while all the other two-mode parameters are as given in the previous section.
 The evolution in $z$ at different $t$ for the  four different cases  of Sec.\,\ref{sec:stab}
is shown in Figs.\,\ref{fig:stab_num}, \ref{fig:nontransp_num}, \ref{fig:conv_num}, and \ref{fig:abs_num}. 
In each case the fastest mode ($v_1$) is represented
in red, while the slowest ($v_2$) is in blue. The two wave packets act as seeds of perturbations, which will generally propagate
from some $(z,\,t)=(z_0,\,t_0)$. The relevant coordinates, especially for unstable (growing) modes are then $z-z_0$ and $t-t_0$. In particular,
rays are defined by constant values of $V=(z-z_0)/(t-t_0)$.

In Fig.\,\ref{fig:stab_num} (completely stable case), the two packets travels in the same direction ($v_1v_2>0$) 
with positive coupling $(\varepsilon>0)$. Besides the wavepackets,
small perturbation are created, and all signals propagate without any amplification. 
In Fig.\,\ref{fig:nontransp_num} (damped case), the two packets travel in opposite directions ($v_1v_2 <0$) with
positive coupling ($\varepsilon>0$). {Perturbations are created in this case too, but are not amplified; actually they are damped, but the size and the numerical observation time are too small to make the damping graphically evident.}
In Fig.\,\ref{fig:conv_num} (convective instability) the two packets travel in the same direction ($v_1v_2 >0$)
with negative coupling. A disturbance is created at some finite $z_0$ and $t_0$, and then grows exponentially but also moves away. We have numerically verified
that the growth of the disturbance closely follows Eq.\,(\ref{eq:growth}) along any ray $V=(z-z_0)/(t-t_0)$ and, 
in particular, that the perturbation is contained in the ray interval
$V\in [v_1,\,v_2]$, and is maximally enhanced along the intermediate
ray $V=(v_1+v_2)/2$ by an exponential growth factor $\propto \varepsilon^{1/2}t$. 
Finally in Fig.\,\ref{fig:abs_num} (absolute instability) the two packets travel in opposite directions ($v_1v_2 <0$) with negative coupling
($\varepsilon<0$). The growth of the disturbance embraces the original point $z_0$ where it has been generated, with the predicted local 
amplification factor---see the comments after Eq.\,(\ref{eq:growth})--(\ref{eq:delta}).

%%%%%%%%%%%%%%%%%%%%%%%%%%%%%%%%%%%%%
\section{Summary and perspectives}
%%%%%%%%%%%%%%%%%%%%%%%%%%%%%%%%%%%%%%%%%%%%%%%%%%%%55

In our work we have discussed in detail a classification of the fast instabilities 
which may arise in the flavor evolution in space-time $(z,t)$ of  
a dense and self-interacting neutrino gas, such as close to the neutrino sphere in core-collapse supernovae.  
This classification is based on
the dispersion relation $D(\omega,\,k)$ among the among the $(t,\,z)$-conjugate coordinates  ($\omega,\,k$).
The dispersion relation has been recently introduced 
 in\,\cite{Izaguirre:2016gsx} for SNe and independently 
elaborated in more general contexts, such as plasma physics and fluid dynamics\,\cite{landau}. 
If the disturbances in the mean field of the $\nu_e\nu_x$  flavor coherence grow, propagating away from the point of origin, they are associated
to  convective instabilities. Conversely, if the disturbances grow in amplitude and extent, embracing the  point of origin, they are called
absolute instabilities. Cases with no instabilities cases may also be differentiated into completely stable (with neither growth nor decay of disturbances)
and damped stable (with decay of disturbances). 

Starting from the dispersion relation, at least for simple systems, one finds that if $\omega$ is real for all real $k$ and vice versa the system is completely stable.
If $\omega$ is real for all real $k$, but $k$ is complex for some real $\omega$, the flow is both stable and damped.
Instead, if $k$ is complex for some real $\omega$ and $\omega$ is also complex for some real $k$, a convective instability arises.
Finally, if $k$ is real for all real $\omega$, and $\omega$ is complex for some real $k$, the instability is absolute.
Deeper criteria can be envisaged to identify instabilities via poles of $D(\omega,\,k)=0$ in the complex $k$ and $\omega$ planes.
In particular,  instabilities can emerge from an evaluation of the time-asymptotic
 behavior of the Green's function of the system, which is related to the dispersion relation by a double (Laplace-Fourier) integral representation,
 amenable to complex-calculus techniques. These techniques, extensively developed in the field of
 plasma physics and fluid dynamics, have been presented and discussed herein in the context of self-interacting neutrinos, with particular 
attention to a simple two-modes system. For a two-beam model of dense neutrinos
 that are  forward-scattering off each other, we considered four cases leading to the four possibilities described above, and presented
a comparison of the predictions from the linear instability theory with the numerical solutions of the linearized equations of motion.
The comparison demonstrate that the two-beam system can be fully understood theoretically and is under control numerically.

The results obtained for the simple two-beam neutrino  model represent a basis to attempt extensions  to more general angular spectra
$G_{\bf v}$ [see Eq.\,(\ref{eq:eln})], as expected in a realistic SN. 
As pointed out in\,\cite{Izaguirre:2016gsx}, one needs a crossing from positive to negative 
$G_{\bf v}$ (i.e., in the ELN) in order to have an instability. Conversely, an ELN with no crossing
would give either a completely stable evolution (if $v_1v_2>0$) or at most a damped stable one (if $v_1v_2<0$). 
The ELN giving fast flavor conversions in SNe would correspond to situations in which there is a significant backward-going neutrino flux, 
corresponding to  $v_1 v_2<0$
in the simple two-beam model. Depending on a possible crossing in the ELN this would lead to damped solutions or to absolute instabilities. 
In this context, 
 a dedicated investigation of  the energy and angle distributions of the neutrino radiation field  has been presented in\,\cite{Tamborra:2017ubu}
 for several spherically symmetric (1D) supernova simulations.
In the cases studied, the ELN near the neutrino-sphere has backward going modes but still does not show any crossing.
 According to our analysis, in first approximation this situation would correspond to the damped stable case, 
 so that no instability (i.e., fast conversion) should  show up in such 1D SN models. However, one cannot exclude
 that these finding may change  in  3D  models,  for  example
in  the  presence  of  LESA  (Lepton-Emission  Self-sustained
Asymmetry)\,\cite{Tamborra:2014aua}. LESA manifests itself in
a pronounced large-scale dipolar pattern in the ELN emission and naturally implies a change of sign in $\nu_e-{\bar\nu}_e$ angular distributions.
 It is therefore conceivable that, especially in the regions where the ELN
changes its sign, crossings in the ELN angular distributions
may occur.  In this case one could expect the emergence of absolute instabilities. In this situation,
the time evolution of the neutrino gas would be dominated by the spectrum of wave-numbers $g_k$ in a given region of space
close to neutrino emission. 
At this regard,  fast flavor conversions have been shown to arise for neutrino angular distributions
inspired by LESA models, assuming space homogeneity \cite{Dasgupta:2016dbv}. 
The next logical step would be to remove the homogeneity assumption, thus introducing a spectrum $g_k$. 
A solution of this problem could be obtained by following the strategy developed in\,\cite{Mangano:2014zda,Mirizzi:2015fva}, i.e., 
by Fourier transforming the equations of motion with respect to the space coordinate, in terms of coupled equations for different conjugate modes. 
If the $k=0$ (homogeneous) mode is unstable, the instability can then cascade to smaller scales by the coupling among the different modes. 

It has also been recently pointed out that a crossing in ELN could be a generic feature of neutrino emission 
from binary neutron-star mergers\,\cite{Wu:2017qpc}. In this case fast conversions would occur, and the associated instability
would be of absolute type. All these cases deserve dedicated studies to gain both a deeper understanding and a broader perspective. 
Indeed, the analysis of the dispersion relation and the relative classification
of the instabilities should be extended to cases presenting many emission angles, both in the zenith and azimuthal ones. 
As pointed out in \cite{Izaguirre:2016gsx}, this would lead to instabilities breaking the axial symmetries, that 
definitely need further investigations.

The use of the dispersion relation to classify the neutrino flavor instabilities could also be extended 
to the study of slow self-induced flavor conversions in SNe. 
In this case the instabilities would depend also on the vacuum  oscillation term. 
The slow self-induced conversions  
occuring for  free streaming neutrinos far from the neutrino-sphere are
dominated by forward modes (i.e., by half-isotropic zenith angle distributions\,\cite{EstebanPretel:2007ec}). 
In the two-beam models, this case would correspond to 
 $v_1 v_2>0$, leading to either a completely stable case or a convective instability. 
Numerical simulations have been found a transition from a stable situation at low-radii (where the neutrino system would exhibit synchronized
oscillations\,\cite{EstebanPretel:2007ec}), to an unstable region at larger radii.
The boundary between these two regions, called synchronization radius, has been taken as a sort
of inner boundary for the subsequent evolution. Using this approach, spatially growing solutions have been found.
Recently,  the assumption of a stationary emission from the spatial  boundary has been 
questioned\,\cite{Abbar:2015fwa,Dasgupta:2015iia,Capozzi:2016oyk}, on the basis that any frequency would
have some amplitude at the boundary that would be amplified in the further evolution. 
In this regard it may happen that the spatially growing solution found in previous works are not stationary.
Indeed, in the presence of convective instability, the perturbations would propagate while they grow. This effect could shift the onset
of the flavor conversions towards larger radii, where the neutrino density is smaller and thus the instability is weaker.
This important issue also requires dedicated investigations.
 
From this discussion it appears that the phenomenology of self-induced flavor conversions in SNe would be much richer than previously thought.
In the presence of fast conversion, the usual characterization of flavor oscillations 
in terms of spatial evolution from an inner boundary should be probably revised, in favor of a time evolution within a small
region of space near the neutrino decoupling region. This would make the approach to SN somewhat similar to the study of neutrinos in the
early Universe. From present studies\,\cite{Sawyer:2005jk,Sawyer:2008zs,Sawyer:2015dsa,Dasgupta:2016dbv}, one might argue
that fast conversions could lead to a quick flavor equilibration among different neutrino species, if instabilities are general enough. 
If flavor equilibration were complete, further oscillation effects
would be ineffective. Otherwise, one could characterize different regimes, e.g., fast conversions near SN core 
followed by spatial slow conversions at larger distances. 
We believe that our current study can provide valuable and inter-disciplinary tools 
to predict the possible scenarios occurring  of the dense SN neutrino gas. The many open questions call for further  numerical
and analytical research in this field.

%%%%%%%%%%%%%%%%%%%%%%%%%%%%%%%%%%%%%%%%%%%%%%%%%%%%%%%%%%%%%%%%%%%%%%%%%%
\section*{Acknowledgments}
%%%%%%%%%%%%%%%%%%%%%%%%%%%%%%%%%%%%%%%%%%%%%%%%%%%%%%%%%%%%%%%%%%%%%

We acknowledge useful discussions with Amol Dighe,  Georg Raffelt, G{\"u}nter Sigl and Irene Tamborra during the development of this work. We also thank Georg Raffelt and Irene Tamborra for useful comments on the  manuscript.
Alessandro Mirizzi acknowledges kind hospitality at TIFR (Mumbai) and at CERN where part of this work was done. 
The work of Francesco Capozzi is supported by NSF Grant PHY-1404311 to J.F. Beacom.
The work of Basudeb Dasgupta  is partially supported by the Dept. of Science and Technology of the Govt.\,of India through a Ramanujam Fellowship and by the Max-Planck-Gesellschaft through a Max-Planck-Partnergroup. The work of 
Eligio Lisi,  Antonio Marrone and Alessandro Mirizzi,  is supported by the Italian
Istituto Nazionale di Fisica Nucleare (INFN) through the ``Theoretical Astroparticle Physics'' project
and by Ministero dell'Istruzione, Universit\`a e Ricerca (MIUR).

%%%%%%%%%%%%%%%%%%%%%%%%%%%%%%%%%%%%%%%%%%%%%%%%%%%%%%%%%%%%%%%%%%%%%%%%5

\end{document}